\documentclass[twocolumn,prb,amsfonts,amsmath,amssymb,floatfix]{revtex4} 
\usepackage{color}
\usepackage{hhline}
\usepackage{mathrsfs}
\usepackage{graphicx}
\usepackage{dcolumn}
\usepackage{bm}
\usepackage{multirow}
\usepackage{booktabs}
\usepackage{afterpage}
\usepackage{amsmath}

\begin{document}

\title{Effective interaction for vanadium oxyhydrides Sr$_{n+1}$V$_n$O$_{2n+1}$H$_n$ ($n=1$ and $n\to \infty$): A constrained-RPA study}

\author{Masayuki Ochi}
\author{Kazuhiko Kuroki}
\affiliation{Department of Physics, Osaka University, Machikaneyama-cho, Toyonaka, Osaka 560-0043, Japan}

\date{\today}
\begin{abstract}
Transition metal oxides have been one of the central objects in the studies of electron correlation effects because of their rich variety of physical properties mainly depending on the transition metal element. On the other hand, exploiting the anion degrees of freedom is less popular but can be another promising way to control properties of strongly correlated materials.
In particular, oxyhydrides offer a unique playground of strongly correlated low-dimensional electronic structure, where the $s$ orbitals of hydrogen breaks a chemical bond between the cation $t_{2g}$ orbitals.
In this study, we evaluate the effective interaction, i.e., the screened Coulomb interaction parameters in low-energy effective models, for vanadium oxyhydrides Sr$_{n+1}$V$_n$O$_{2n+1}$H$_n$ ($n=1,\infty$) using the constrained random-phase approximation (cRPA).
We find that the effective interaction in the $t_{2g}$ model, where only the $t_{2g}$ orbitals are explicitly considered, is strongly screened by the $e_g$ bands compared with that for oxides, because the $e_g$ bands are much entangled with the $t_{2g}$ bands in the oxyhydrides.
On the other hand, the effective interaction is rather strong in the $d$ model, where all the vanadium $d$ orbitals are explicitly considered, owing to a large energy separation between the V-$d$ bands and the anion bands (O-$p$ and H-$s$), because the O-$p$ states are stabilized by the existence of the hydrogen atoms.
These findings suggest that non-trivial and unique correlation effects can take place in vanadium oxyhydrides.
\end{abstract}

\maketitle

\section{Introduction}
Transition metal oxides are one of the most popular playgrounds for strong correlation effects~\cite{SC}.
For example, transition metal oxides with the Ruddlesden-Popper (RP) phase, $A_{n+1}B_n$O$_{3n+1}$ ($n=1,2,\dots,\infty$) with $B$ being a transition metal element, have a very simple layered crystal structure but exhibit several intriguing properties such as unconventional superconductivity in cuprates~\cite{cup}.
Physical properties of transition metal oxides are dominated mainly by the transition metal element.
In addition, changing the $A$ site element often alters materials properties, e.g., by the chemical pressure effect through the difference of its atomic radius, which sometimes induces a structural transition, and by the carrier doping effect through the difference of the valence number among $A$ site elements, such as Sr$^{2+}$ and La$^{3+}$.

An anion is another degree of freedom to control materials properties in transition metal oxides.
For example, some kinds of cuprate superconductors with multiple anions, such as La$_2$CuO$_4$F$_x$~\cite{LaCuOF}, Nd$_2$CuO$_{4-x}$F$_y$~\cite{NdCuOF}, Sr$_2$CuO$_2$F$_{2+\delta}$~\cite{SrCuOF}, and (Ca$_{1-x}$Na$_x$)$_2$CuO$_2$Cl$_2$~\cite{CuOCl}, exhibit a superconducting transition at several tens of Kelvin.
In these materials, fluorine or chlorine doping changes not only the carrier concentration but also the local environment around copper, which yields a different crystal field from oxides.
Intercalated anions, e.g., fluorine atoms in Sr$_3$Ru$_2$O$_7$F$_2$~\cite{Ru3272}, can reduce the three-dimensionality in layered structures by separating the layers along the stacked direction.
Such compounds with multiple anions, named mixed-anion compounds, have recently attracted much attention owing to their possibilities of realizing novel functionalities in a different way from oxides~\cite{mixed_review}.

In particular, among mixed-anion compounds, oxyhydrides are materials with unique and remarkable aspects because of the distinctive nature of hydrogen.
For example, heavy electron doping enabled by hydrogen revealed two-dome superconducting phases neighboring with two different types of antiferromagnetic phases in LaFeAsO$_{1-x}$H$_x$~\cite{FeAs1,FeAs2}.
It is remarkable that several transition metal oxyhydrides have been reported in very recent years~\cite{TiOH1,TiOH2,TiOH3,TiOH4,CoOH1,CoOH2,CrOH,MgOH,ScOH,TiOH5,CoOH3,BaVOH}.
In vanadium oxyhydrides Sr$_{n+1}$V$_n$O$_{2n+1}$H$_n$ ($n=1, 2, \infty$)~\cite{SVOH_Angew,SVOH_JACS},
it was pointed out that chemical bonds among the V-$t_{2g}$ orbitals through the O-$p$ orbitals are partially lost when oxygen is partially replaced with hydrogen, because the H-$s$ orbital has a different parity from the V-$t_{2g}$ orbitals.
Because hydrogen atoms are aligned in vanadium oxyhydrides~\cite{SVOH_Angew,SVOH_JACS}, this role called a $\pi$-blocker\cite{SVOH_Yamamoto} decreases the dimensionality of the electronic structure.
These studies also pointed out that the symmetry of the crystal field around vanadium is lowered by hydrogen.

\begin{figure}
\begin{center}
\includegraphics[width=7 cm]{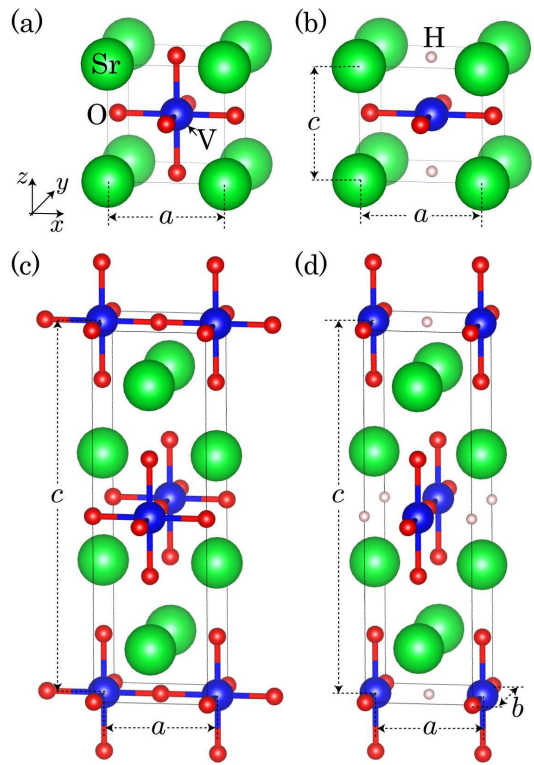}
\caption{Crystal structure of (a) SrVO$_3$, (b) SrVO$_2$H, (c) Sr$_2$VO$_4$, and (d) Sr$_2$VO$_3$H depicted using the VESTA software~\cite{VESTA}.}
\label{fig:crys}
\end{center}
\end{figure}

Because Sr$_{n+1}$V$_n$O$_{3n+1}$ ($n=1,\infty$) (Fig.~\ref{fig:crys}(a) for $n=\infty$ and \ref{fig:crys}(c) for $n=1$) have been a text-book compound for theoretical investigation of the electron correlation effects (e.g., Ref.~\onlinecite{Silke}),
it is important to study the electronic structure of the corresponding oxyhydrides Sr$_{n+1}$V$_n$O$_{2n+1}$H$_n$ ($n=1,\infty$) (Fig.~\ref{fig:crys}(b)(d)).
This importance is also supported from experimental studies revealing that Sr$_{n+1}$V$_n$O$_{2n+1}$H$_n$ are strongly correlated materials.
For example, an antiferromagnetic order with an anomalously reduced magnetic moment was observed for $n=1,2,\infty$~\cite{SVOH_Angew}.
While the insulating state is realized at ambient pressure for $n=\infty$~\cite{SVOH_epitaxial}, a metal-insulator transition is induced by applying pressure~\cite{SVOH_Yamamoto}.
Although some studies reported first-principles electronic structure of Sr$_{n+1}$V$_n$O$_{2n+1}$H$_n$ calculated using density functional theory (DFT) and discussed their magnetic properties~\cite{SVOH_Yamamoto,SVOH_theory,SVOH_theory_wan}, more elaborate theoretical treatment of correlation effects is often required for strongly correlated materials.
For this purpose, it is essential to construct the model Hamiltonian representing the low-energy electronic structure, including the evaluation of the effective Coulomb interaction parameters, by first-principles calculation~\cite{downfold}.
However, first-principles evaluation of such parameters for oxyhydrides has still been missing.
We note that, although the magnetic interaction parameters calculated by the DFT$+U$ method as presented in Refs.~\onlinecite{SVOH_theory} and \onlinecite{SVOH_theory_wan} are helpful for understanding the anisotropy of the magnetic interaction (i.e., when discussing their relative strength) in oxyhydrides, it is problematic that these parameters can vary by changing the $U$ parameter assumed in the DFT+$U$ calculations, in addition to the fact that the magnetic interaction is evaluated at the DFT level there.

In this study, we evaluate the screened Coulomb interaction parameters in low-energy effective models for Sr$_{n+1}$V$_n$O$_{2n+1}$H$_n$ ($n=1,\infty$) using the constrained random-phase approximation (cRPA)~\cite{crpa}.
For this purpose, we start from the DFT band structure and verify the low-dimensional electronic structure in these materials as previous studies pointed out.
As for the interaction parameters evaluated by cRPA, we find that the effective interaction in the V-$t_{2g}$ model, where only the $t_{2g}$ orbitals are explicitly considered, is sizably screened by the V-$e_g$ bands because of strong entanglement between the $t_{2g}$ and $e_g$ bands.
On the other hand, for the V-$d$ model, where all the V-$d$ orbitals are explicitly considered, the effective interaction is stronger than that for the oxides because of a large energy separation between the V-$d$ bands and the anion bands (O-$p$ and H-$s$).
These findings suggest that possibly non-trivial and unique correlation effects can be realized in vanadium oxyhydrides, and also that special care must be taken in choosing which model to adopt in order to analyze the low energy properties.

This paper is organized as follows.
Section~\ref{sec:method} presents a brief overview of the cRPA formulation, and some computational conditions are shown in Sec.~\ref{sec:detail}.
Sections~\ref{sec:results113} and \ref{sec:results214} present our calculation results for $n=\infty$ and $n=1$ compounds, respectively.
Our findings are summarized in Sec.~\ref{sec:sum}.

\section{Method\label{sec:method}}

We briefly review the formulation of cRPA, which was used to evaluate the interaction parameters of the low-energy effective models in our study.
Because we concentrate on the static interaction, we show the cRPA formulation only for the static variables.

One begins with the Kohn-Sham orbitals $\phi_{kn}$ and their eigenvalues $\epsilon_{kn}$, where $k = ({\bf k},\sigma)$ is a combined index for the $k$-vector and the spin $\sigma$ and $n$ is the band index.
Then, the static independent-particle polarization function reads
\begin{align}
\chi_0({\bf r},{\bf r'}) &= \sum_{kn}^{\mathrm{occ.}} \sum_{k' n'}^{\mathrm{unocc.}} \frac{1}{\epsilon_{kn} - \epsilon_{k'n'}} \notag \\
&\times(\phi_{kn}^*(r) \phi_{k'n'}(r) \phi_{k'n'}^*(r') \phi_{kn}(r')\notag \\
&+\phi_{kn}^*(r') \phi_{k'n'}(r') \phi_{k'n'}^*(r) \phi_{kn}(r)),
\end{align}
where $kn$ and $k'n'$ are the indices of the occupied and unoccupied Kohn-Sham orbitals, respectively.
In cRPA, one should exclude the electron excitations within the correlated subspace spanned by the Wannier orbitals (see Ref.~\onlinecite{cRPA_entangle} for more details about the treatment of the band entanglement).
By denoting the rest of the polarization function as $\chi^r_0({\bf r},{\bf r'})$, the dielectric function $\epsilon$ in cRPA reads
\begin{equation}
\epsilon = 1 - v \chi_0^r,
\end{equation}
where $v$ is the bare Coulomb interaction. Finally, we obtain the screened Coulomb interaction,
\begin{equation}
W = \epsilon^{-1} v.
\end{equation}
By using $W$, the effective interaction parameters between the Wannier functions $\psi_n({\bf r})$ and $\psi_m({\bf r})$ are evaluated as follows:
\begin{align}
U^{\mathrm{scr}}_{nm} = \int d{\bf r} d{\bf r'} |\psi_n({\bf r})|^2 W({\bf r},{\bf r'}) |\psi_m({\bf r'})|^2,\\
J^{\mathrm{scr}}_{nm} = \int d{\bf r} d{\bf r'} \psi_n^*({\bf r})^2 \psi_m({\bf r}) W({\bf r},{\bf r'}) \psi_n({\bf r'}) \psi^*_m({\bf r'}),
\end{align}
for the direct Coulomb and exchange interactions, respectively. When the screened interaction $W^r$ in the above integrals is replaced with the bare interaction $v$,
we shall denote these variables as $U^{\mathrm{bare}}_{nm}$ and $J^{\mathrm{bare}}_{nm}$, respectively.

\section{Computational details\label{sec:detail}}

First, we calculated the first-principles band structure using the \textsc{Quantum ESPRESSO} code~\cite{QE1,QE2}.
Perdew-Burke-Ernzerhof parametrization of the generalized gradient approximation (PBE-GGA)~\cite{PBE}
and the scalar-relativistic version of the optimized norm-conserving Vanderbilt pseudopotentials~\cite{pp_opt} 
taken from PseudoDojo~\cite{pseudodojo} were used.
For the pseudopotentials, core electrons of each element are as follows: [He] for O, [Ne] for V and Cr, and [Ar]$3d^{10}$ for Sr (i.e., V-$3s^23p^6$ and Sr-$4s^24p^6$ semicore states are treated as valence).
Experimental crystal structures were taken from Ref.~\onlinecite{SVO3_strct} for SrVO$_3$, Ref.~\onlinecite{SVOH_Angew} (data taken at 5 K) for SrVO$_2$H, Ref.~\onlinecite{SVOH_JACS} for Sr$_2$VO$_4$ and Sr$_2$VO$_3$H, and Ref.~\onlinecite{SCrO3_strct} for SrCrO$_3$.
The plane-wave cutoff energy of 150 Ry, a $12\times12\times12$ $k$-mesh for SrVO$_3$, SrVO$_2$H, and SrCrO$_3$, and a $10\times10\times10$ $k$-mesh for Sr$_2$VO$_4$ and Sr$_2$VO$_3$H, were used with the Gaussian smearing width of 0.02 Ry.

Next, we extracted (maximally localized) Wannier functions~\cite{wannier1,wannier2} using the RESPACK code~\cite{resp1,resp2,resp3,resp4,resp5}, by which we also obtained the hopping parameters among the Wannier functions.
Finally, we evaluated the interaction parameters among the Wannier functions using cRPA~\cite{crpa} as implemented in the RESPACK code.
For this purpose, the cutoff energy of the dielectric function was set to 40 Ry for all the compounds. The total number of bands (i.e., the sum of the numbers of the valence and conduction bands)
considered in our cRPA calculation was 200 for SrVO$_3$, SrVO$_2$H, and SrCrO$_3$, and 400 for Sr$_2$VO$_4$ and Sr$_2$VO$_3$H, unless noted.

In this paper, the $t_{2g}$, $d$, $dp$, and $dps$ models denote the low-energy effective models consisting of the V(Cr)-$d_{xy,yz,xz}$, V(Cr)-$d$, V-$d$ $+$ O-$p$, V-$d$ $+$ O-$p$ $+$ H-$s$ orbitals, respectively.
Although $t_{2g}$ is an inappropriate name for oxyhydrides with a lowered crystal-field symmetry in the strict sense of the term, we call the $d_{xy,yz,xz}$ orbitals the `$t_{2g}$ orbitals' for simplicity. We also call the remaining $d$ orbitals the `$e_g$ orbitals'.

\section{Results and discussions}

For all the compounds investigated in this study, we shall show their hopping and interaction parameters only partially in the main text.
A more extensive list of these parameters is shown in Appendices A--B.

\subsection{SrVO$_3$ and SrVO$_2$H ($n=\infty$)\label{sec:results113}}

\subsubsection{Band structure and Wannier functions}

\begin{figure}
\begin{center}
\includegraphics[width=8.5 cm]{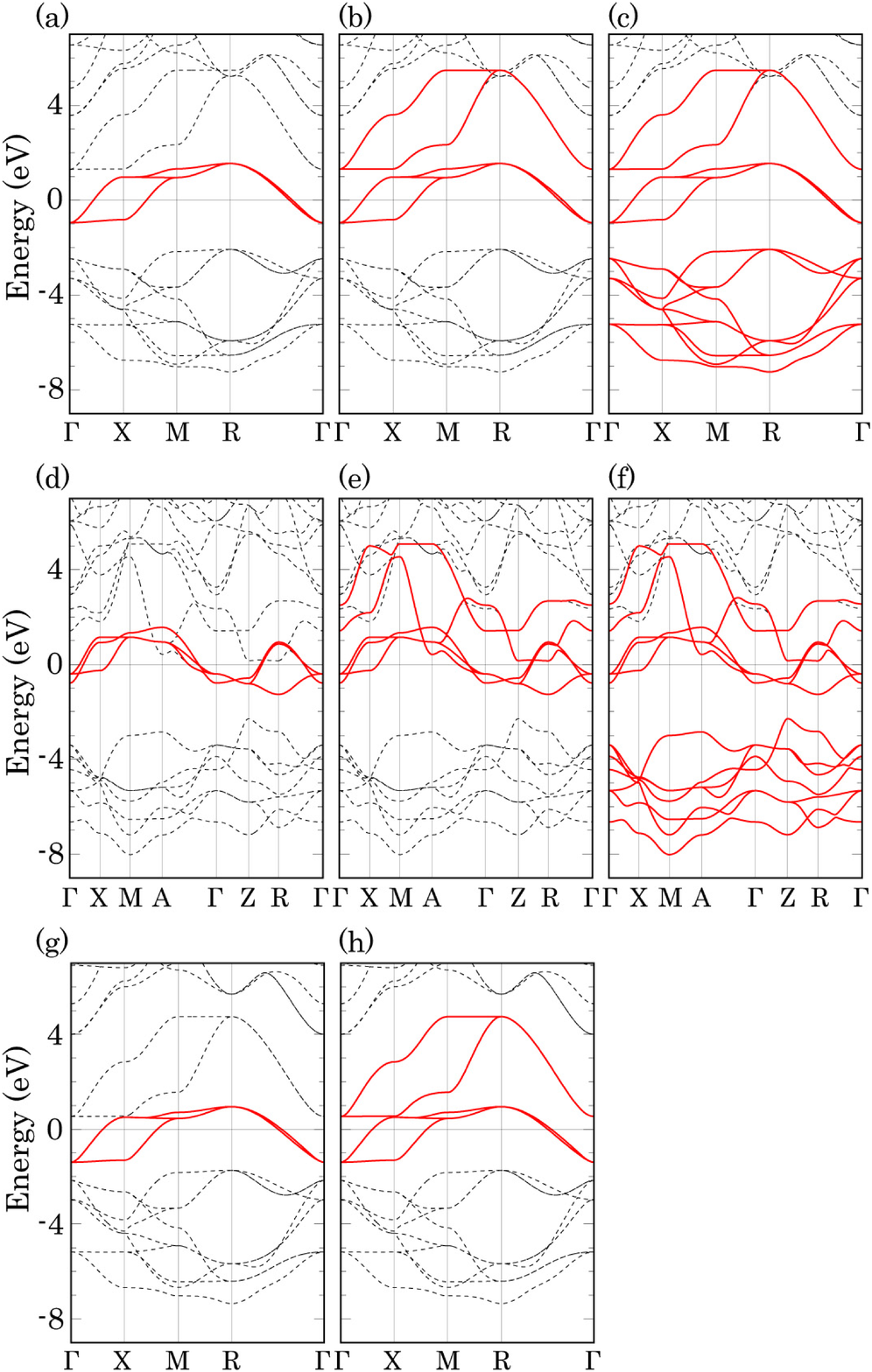}
\caption{Calculated electronic band structure of (a)--(c) SrVO$_3$, (d)--(f) SrVO$_2$H, and (g)--(h) SrCrO$_3$. First-principles band structure is shown with black broken lines and the band dispersion calculated with the tight-binding model consisting of the Wannier functions is shown with red solid lines. Corresponding effective models are $t_{2g}$ for panels (a)(d)(g), $d$ for panels (b)(e)(h), $dp$ for panel (c), and $dps$ for panel (f).}
\label{fig:band1}
\end{center}
\end{figure}

\begin{figure}
\begin{center}
\includegraphics[width=8.5 cm]{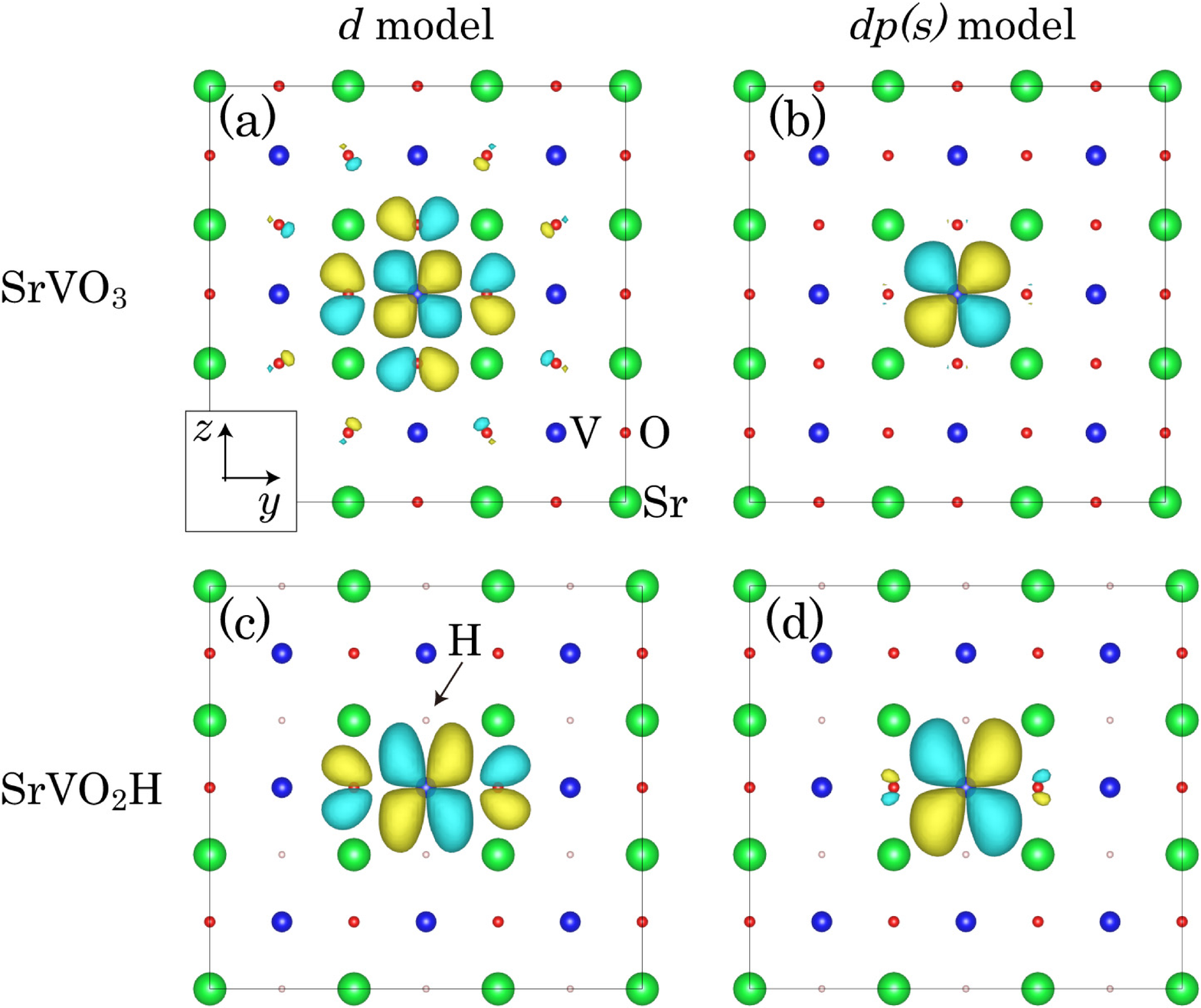}
\caption{Wannier orbitals of V-$d_{yz}$ for (a)--(b) SrVO$_3$ and (c)--(d) SrVO$_2$H.
Corresponding effective models are $d$ for panels (a)(c), $dp$ for panel (b), and $dps$ for panel (d).}
\label{fig:wan_dyz}
\end{center}
\end{figure}

\begin{figure}
\begin{center}
\includegraphics[width=8.5 cm]{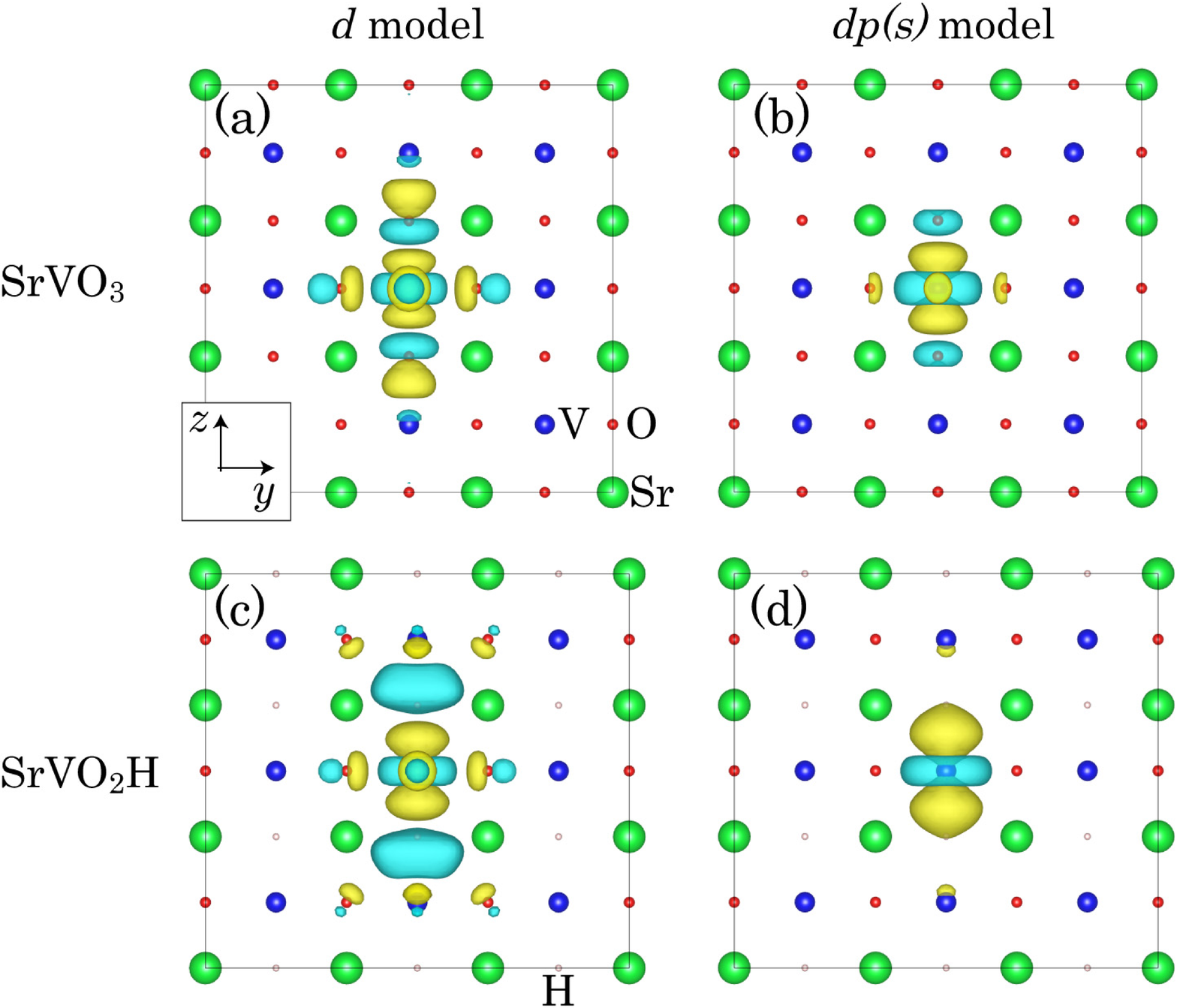}
\caption{Wannier orbitals of V-$d_{3z^2-r^2}$ for (a)--(b) SrVO$_3$ and (c)--(d) SrVO$_2$H.
Corresponding effective models are $d$ for panels (a)(c), $dp$ for panel (b), and $dps$ for panel (d).}
\label{fig:wan_dz2}
\end{center}
\end{figure}

\begin{figure}
\begin{center}
\includegraphics[width=7.5 cm]{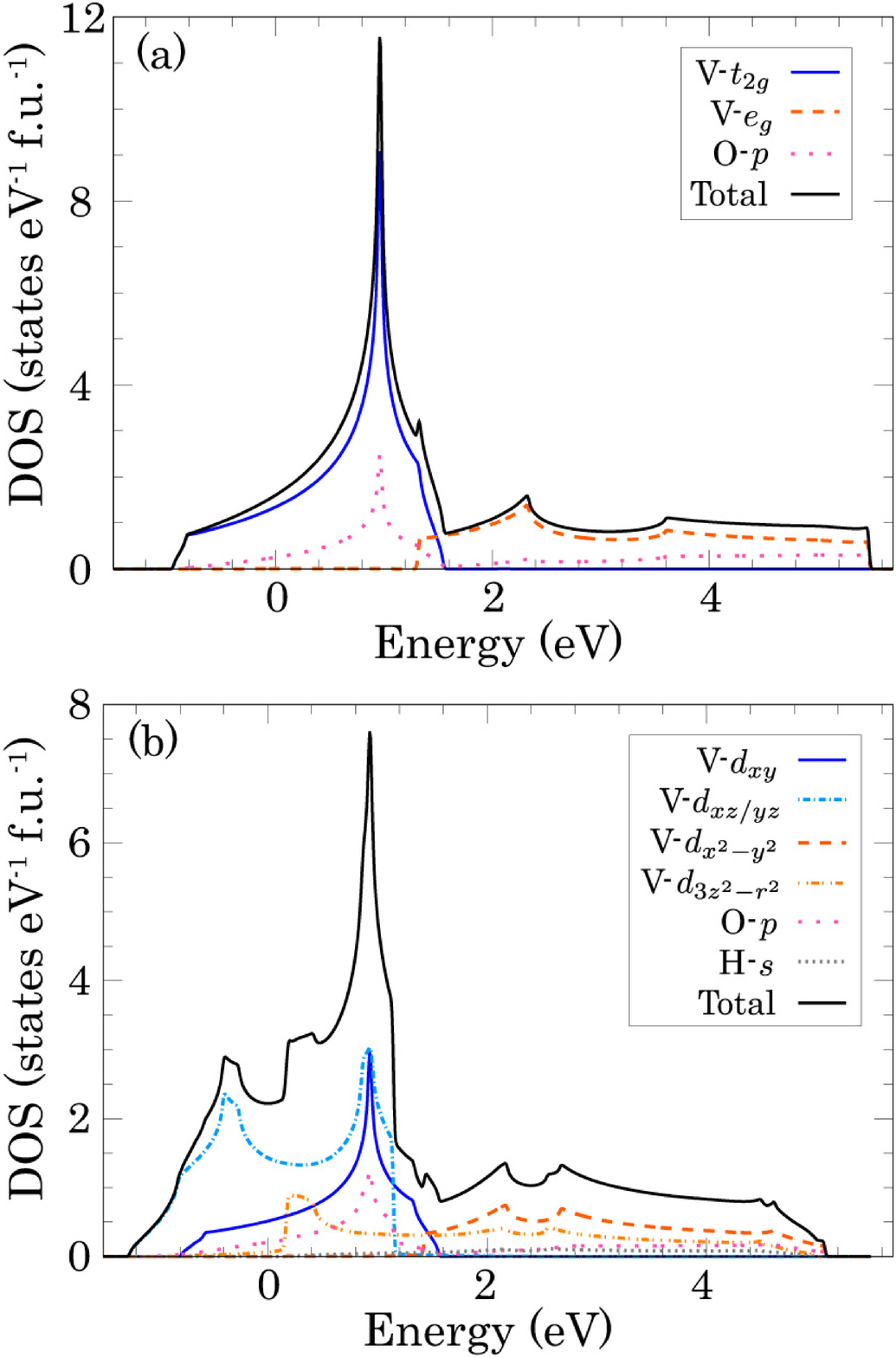}
\caption{(Partial) DOS for (a) SrVO$_3$ and (b) SrVO$_2$H calculated with our $dp(s)$ tight-binding model.}
\label{fig:113_dos}
\end{center}
\end{figure}

Figure~\ref{fig:band1} presents the calculated electronic band structure of SrVO$_3$, SrVO$_2$H, and SrCrO$_3$.
Here, we calculated the electronic structure of SrVO$_3$ to compare it with that for SrVO$_2$H.
Because SrVO$_3$ and SrVO$_2$H have different $d$-electron occupation numbers, $d^1$ for the former and $d^2$ for the latter, we also show some results for SrCrO$_3$ with $d^2$ configuration to enable more detailed comparison among them.
In Fig.~\ref{fig:band1}, the band structure calculated with the tight-binding model consisting of the Wannier functions are shown with red solid lines along with the first-principles one with black broken lines.
The corresponding tight-binding models are the $t_{2g}$ model in Fig.~\ref{fig:band1}(a)(d)(g), the $d$ model in Fig.~\ref{fig:band1}(b)(e)(h), the $dp$ model in Fig.~\ref{fig:band1}(c), and the $dps$ model in Fig.~\ref{fig:band1}(f).

The band structure of SrVO$_2$H is similar to but in part different from those for SrVO$_3$ and SrCrO$_3$.
For example, the $t_{2g}$ band dispersion along the $\Gamma$-X-M-A-$\Gamma$ line in SrVO$_2$H, shown with red solid lines in Fig.~\ref{fig:band1}(b), is very similar to that along the $\Gamma$-X-M-R-$\Gamma$ line in the oxides, shown with red solid lines in Fig.~\ref{fig:band1}(a)(g).
We note that both these two $k$-paths represent $(0,0,0)$-$(\pi /a,0,0)$-$(\pi /a, \pi /a, 0)$-$(\pi /a, \pi /a, \pi /c)$-$(0,0,0)$ in the Cartesian coordinate, where $a$ and $c$ ($=a$ for the oxides) are the lattice constants shown in Fig.~\ref{fig:crys}(a)--(b).
On the other hand, the $t_{2g}$ bands show a small dispersion along the $k_z$ direction, such as along the $\Gamma$-Z line, for SrVO$_2$H, unlike the corresponding band dispersion in the oxides, i.e., those along the $\Gamma$-X line.
Such a small band dispersion along the $k_z$ direction is induced by the hydrogen atom placing along the $z$ direction as shown in Fig.~\ref{fig:crys}(b).
In other words, the H-$s$ orbital cannot form a chemical bond with the V-$t_{2g}$ orbitals because of their different parities~\cite{SVOH_Angew,SVOH_JACS}.
Such a low-dimensionality is characteristic of oxyhydrides.

To see the low-dimensionality of the $t_{2g}$ states in more detail, we depicted the $d_{yz}$ Wannier orbitals in Fig.~\ref{fig:wan_dyz}.
As is consistent with the previous theoretical study~\cite{SVOH_theory_wan}, the $t_{2g}$ Wannier orbital in the $t_{2g}$ or $d$ models (the former not shown here), which can be usually regarded as an anti-bonding pair of the atomic orbitals of the cation and the surrounding anions, has no weight on hydrogen sites.
It is also noteworthy that the $t_{2g}$ orbital tends to extend in SrVO$_2$H~\cite{SVOH_theory_wan}. This feature is maintained also in the $dp(s)$ model as shown in Fig.~\ref{fig:wan_dyz}(b)(d), suggesting that this delocalization is partially brought by a lower-energy crystal field in SrVO$_2$H, where O$^{2-}$ is partially replaced with H$^{-}$.

Such a low-dimensionality can also be seen in Fig.~\ref{fig:113_dos}.
While a shape of the density of states (DOS) characteristic of two-dimensional electronic structure on the square lattice can be seen for the $t_{2g}$ orbitals in SrVO$_3$ and the $d_{xy}$ orbital in SrVO$_2$H, a strong DOS enhancement near the band edge, which is characteristic of (quasi-)one-dimensional electronic structure, is realized for the $d_{xz/yz}$ orbitals in SrVO$_2$H.

While we mainly focused on the $t_{2g}$ orbitals so far, the $e_g$ orbitals in SrVO$_2$H, which can form a chemical bond with the H-$s$ orbital as shown in Fig.~\ref{fig:wan_dz2}(c), are also quite different from those in SrVO$_3$.
For example, as shown in Fig.~\ref{fig:band1}(e), the energy levels of the $e_g$ bands in SrVO$_2$H are much lowered by hydrogen compared with SrVO$_3$.
As a result, the bottom of the $e_g$ bands is very close to the Fermi energy in SrVO$_2$H, which can also be seen in Fig.~\ref{fig:113_dos}(b).
We shall come back to this point later in this paper.

\subsubsection{Hopping parameters}

\begin{table}
\begin{center}
\begin{tabular}{c c c c c c c c}
\hline \hline
 & & $t_x$ & $t_y$ & $t_z$ & $\Delta$ & $U^{\mathrm{scr}}_{t_{2g}}$ & $U^{\mathrm{bare}}_{t_{2g}}$  \\
\hline
SrVO$_3$ & $d_{xy}$ & $-0.26$ & $-0.26$ & $-0.03$ & - & 3.42 & 15.78\\
 & & & & & & 3.48 [\onlinecite{resp1}] &  \\
 & & & & & & 3.2 [\onlinecite{Vaugier_crpa}] & 16.1 [\onlinecite{Vaugier_crpa}] \\
  & & & & & & 3.39 [\onlinecite{nomura_crpa}] & 15.0 [\onlinecite{nomura_crpa}] \\
  & & & & & & 3.36 [\onlinecite{nomura_crpa}] & 16.0 [\onlinecite{nomura_crpa}] \\
  & & & & & & 3.4 [\onlinecite{miyake_crpa}] &  \\
  & & & & & & 3.3 [\onlinecite{casula_crpa}] &  \\
 \hline
 SrVO$_2$H & $d_{xy}$ & $-0.25$ & $-0.25$ & $-0.04$  & - & 3.00 & 16.04 \\
 & $d_{yz}$ & $0.01$ & $-0.42$ & $0.10$ & $-0.45$ & 2.60 & 15.18\\ 
 \hline
SrCrO$_3$ & $d_{xy}$ & $-0.24$ & $-0.24$ & $-0.02$ & - & 2.97 & 16.18\\
 & & & & & & 2.7 [\onlinecite{Vaugier_crpa}] & 16.4 [\onlinecite{Vaugier_crpa}] \\
\hline \hline
\end{tabular}
\end{center}
\caption{\label{table:113_t2g} Hopping and interaction parameters (in eV) for the $t_{2g}$ model. Equivalent orbitals to the listed ones, e.g., $d_{xz}$ in SrVO$_2$H, are omitted in this table.}
\end{table}

\begin{table}
\begin{center}
\begin{tabular}{c c c c c c c c}
\hline \hline
 & & $t_x$ & $t_y$ & $t_z$ &  $\Delta$ & $U^{\mathrm{scr}}_d$ & $U^{\mathrm{bare}}_d$ \\
\hline
SrVO$_3$ & $d_{xy}$ & $-0.26$ & $-0.26$ & $-0.02$  & - & 3.43 & 15.85 \\
& & & & & & 3.5 [\onlinecite{aryasetiawan_crpa}] &  \\
 & $d_{x^2-y^2}$ & $-0.51$ & $-0.51$ & $0.00$ & 2.76 & 3.57 & 16.36\\
 & $d_{3z^2-r^2}$ & $-0.17$ & $-0.17$ & $-0.67$ & 2.76 & 3.57 & 16.36\\
 \hline
 SrVO$_2$H & $d_{xy}$ & $-0.25$ & $-0.25$ & $-0.04$  & - & 3.97 & 16.06 \\
 & $d_{yz}$ & $0.01$ & $-0.42$ & $0.10$ & $-0.44$ & 3.75 & 15.28 \\
 & $d_{x^2-y^2}$ & $-0.44$ & $-0.44$ & $0.01$ & $2.49$ & 4.04 & 16.37 \\
 & $d_{3z^2-r^2}$ & $-0.09$ & $-0.09$ & $0.88$ & $1.52$ & 3.26 & 13.58\\
  \hline
 SrCrO$_3$ & $d_{xy}$ & $-0.24$ & $-0.24$ & $-0.02$  & - & 3.04 & 16.20 \\
 & $d_{x^2-y^2}$ & $-0.51$ & $-0.51$ & $0.00$  & 2.52 & 3.18 & 16.82 \\
  & $d_{3z^2-r^2}$ & $-0.17$ & $-0.17$ & $-0.68$  & 2.52 & 3.18 & 16.82 \\
\hline \hline
\end{tabular}
\end{center}
\caption{\label{table:113_d} Hopping and interaction parameters (in eV) for the $d$ model. Equivalent orbitals to the listed ones are omitted in this table.}
\end{table}

A portion of the hopping parameters is shown in Tables~\ref{table:113_t2g} and \ref{table:113_d}, where $t_i$ ($i=x,y,z$) denotes the nearest-neighbor hopping parameter along the $i$ direction between the same type of the orbital (e.g., $d_{yz}$-$d_{yz}$), and
$\Delta$ is the on-site energy relative to the $d_{xy}$ orbital.
We note that these parameters are not sufficient to reproduce the first-principles band structure.
We just show them to discuss the dimensionality of the electronic structure.
A set of the hopping parameters for the $t_{2g}$ and $d$ models that can well reproduce the band dispersion are provided in Appendix A.

In Table~\ref{table:113_t2g}, we can see that the hopping parameters for the $d_{xy}$ orbital are almost the same among SrVO$_3$, SrVO$_2$H, and SrCrO$_3$.
On the other hand, hydrogen atoms yield a drastically suppressed value of $t_z$, $-0.04$ eV, for the $d_{xz/yz}$ orbitals in SrVO$_2$H.
As a result, the quasi-one-dimensional electronic structure is realized for the $d_{xz/yz}$ orbitals in SrVO$_2$H as we have seen in the previous section.
In the previous section, we have also mentioned that a sizably increased value of $t_y$ for the $d_{yz}$ orbital ($t_x$ for the $d_{xz}$ orbital) in SrVO$_2$H ($-0.42$ eV) from that in SrVO$_3$ ($-0.26$ eV) is another characteristic feature of oxyhydrides, which was pointed out in the previous theoretical study on SrCrO$_2$H with a hypothetically hydrogen-ordered structure~\cite{SVOH_theory_wan}.
This feature also enhances the low-dimensionality of the $d_{xz/yz}$ states in SrVO$_2$H.
The crystal-field splitting induced by hydrogen can be seen in Table~\ref{table:113_t2g}: the on-site energy of the $d_{xz/yz}$ orbitals relative to the $d_{xy}$ orbital becomes a sizable negative value ($-0.45$ eV) in SrVO$_2$H.
All the features of the hopping parameters for the $t_{2g}$ orbitals mentioned above are maintained also in the $d$ model, as shown in Table~\ref{table:113_d}.

In Table~\ref{table:113_d}, we can see that the $d_{3z^2-r^2}$ orbital in SrVO$_2$H, which has a strong chemical bond with the H-$s$ orbital as shown in Fig.~\ref{fig:wan_dz2}(c), exhibits an enhanced value of $t_z$ (0.88 eV) together with a lowered on-site energy ($\sim 1$ eV lower than that for the $d_{x^2-y^2}$ orbital). The sign of $t_z$ for the $d_{3z^2-r^2}$ orbital is changed from the oxides to oxyhydride, which originates from the different parity of the O-$p_z$ orbital in oxides and the H-$s$ orbital in oxyhydride.

\subsubsection{On-site direct Coulomb interaction}

We next move on to the effective Coulomb interaction parameters obtained by our cRPA calculation.
We start from the $dp(s)$ model because the screening processes taken into account are most limited there.
The screened interaction for the on-site direct Coulomb terms among the V-$d$ orbitals in SrVO$_3$ is
\begin{equation}
U^{\mathrm{scr}}_{dp} = \left(
\begin{array}{ccccc}
11.41 \ \ &\ \  10.01 \ \ &\ \  10.01 \ \ &\ \ 11.02 \ \ &\ \  10.31\\
10.01 \ \ &\ \  11.41 \ \ &\ \  10.01 \ \ &\ \ 10.49 \ \ &\ \  10.84\\
10.01 \ \ &\ \  10.01 \ \ &\ \  11.41 \ \ &\ \ 10.49 \ \ &\ \  10.84\\
11.02 \ \ &\ \  10.49 \ \ &\ \  10.49 \ \ &\ \ 12.64 \ \ &\ \  10.83\\
10.31 \ \ &\ \  10.84 \ \ &\ \  10.84 \ \ &\ \ 10.83 \ \ &\ \  12.64
\end{array}
\right),
\end{equation}
while the bare interaction is
\begin{equation}
U^{\mathrm{bare}}_{dp} = \left(
\begin{array}{ccccc}
19.36 \ \ &\ \  17.72 \ \ &\ \  17.72 \ \ &\ \ 19.28 \ \ &\ \  18.28\\
17.72 \ \ &\ \  19.36 \ \ &\ \  17.72 \ \ &\ \ 18.53 \ \ &\ \  19.03\\
17.72 \ \ &\ \  17.72 \ \ &\ \  19.36 \ \ &\ \ 18.53 \ \ &\ \  19.03\\
19.28 \ \ &\ \  18.53 \ \ &\ \  18.53 \ \ &\ \ 21.31 \ \ &\ \  19.16\\
18.28 \ \ &\ \  19.03 \ \ &\ \  19.03 \ \ &\ \ 19.16 \ \ &\ \  21.31\\
\end{array}
\right),
\end{equation}
where the orbital index runs as $d_{xy}$, $d_{yz}$, $d_{xz}$, $d_{x^2-y^2}$, and $d_{3z^2-r^2}$.
For SrVO$_2$H, we obtained
\begin{equation}
U^{\mathrm{scr}}_{dps} = \left(
\begin{array}{ccccc}
8.16 \ \ &\ \  6.48 \ \ &\ \ 6.48 \ \ &\ \ 7.61 \ \ &\ \  6.53 \\
6.48 \ \ &\ \  7.28 \ \ &\ \ 6.17 \ \ &\ \ 6.72 \ \ &\ \  6.65 \\
6.48 \ \ &\ \  6.17 \ \ &\ \ 7.28 \ \ &\ \ 6.72 \ \ &\ \  6.65 \\
7.61 \ \ &\ \  6.72 \ \ &\ \ 6.72 \ \ &\ \ 8.95 \ \ &\ \  6.79 \\
6.53 \ \ &\ \  6.65 \ \ &\ \ 6.65 \ \ &\ \ 6.79 \ \ &\ \  7.90
\end{array}
\right),
\end{equation}
and
\begin{equation}
U^{\mathrm{bare}}_{dps} = \left(
\begin{array}{ccccc}
18.59 \ \ &\ \  16.32 \ \ &\ \  16.32 \ \ &\ \ 18.57 \ \ &\ \  16.59 \\
16.32 \ \ &\ \  17.02 \ \ &\ \  15.67 \ \ &\ \ 17.06 \ \ &\ \  16.58 \\
16.32 \ \ &\ \  15.67 \ \ &\ \  17.02 \ \ &\ \ 17.06 \ \ &\ \  16.58 \\
18.57 \ \ &\ \  17.06 \ \ &\ \  17.06 \ \ &\ \ 20.55 \ \ &\ \  17.39 \\
16.59 \ \ &\ \  16.58 \ \ &\ \  16.58 \ \ &\ \ 17.39 \ \ &\ \  18.26
\end{array}
\right).
\end{equation}
Here, we omit other matrix elements such as $d$-$p$ interaction, which are shown in Appendices B-1 and B-2.

\begin{figure}
\begin{center}
\includegraphics[width=7.5 cm]{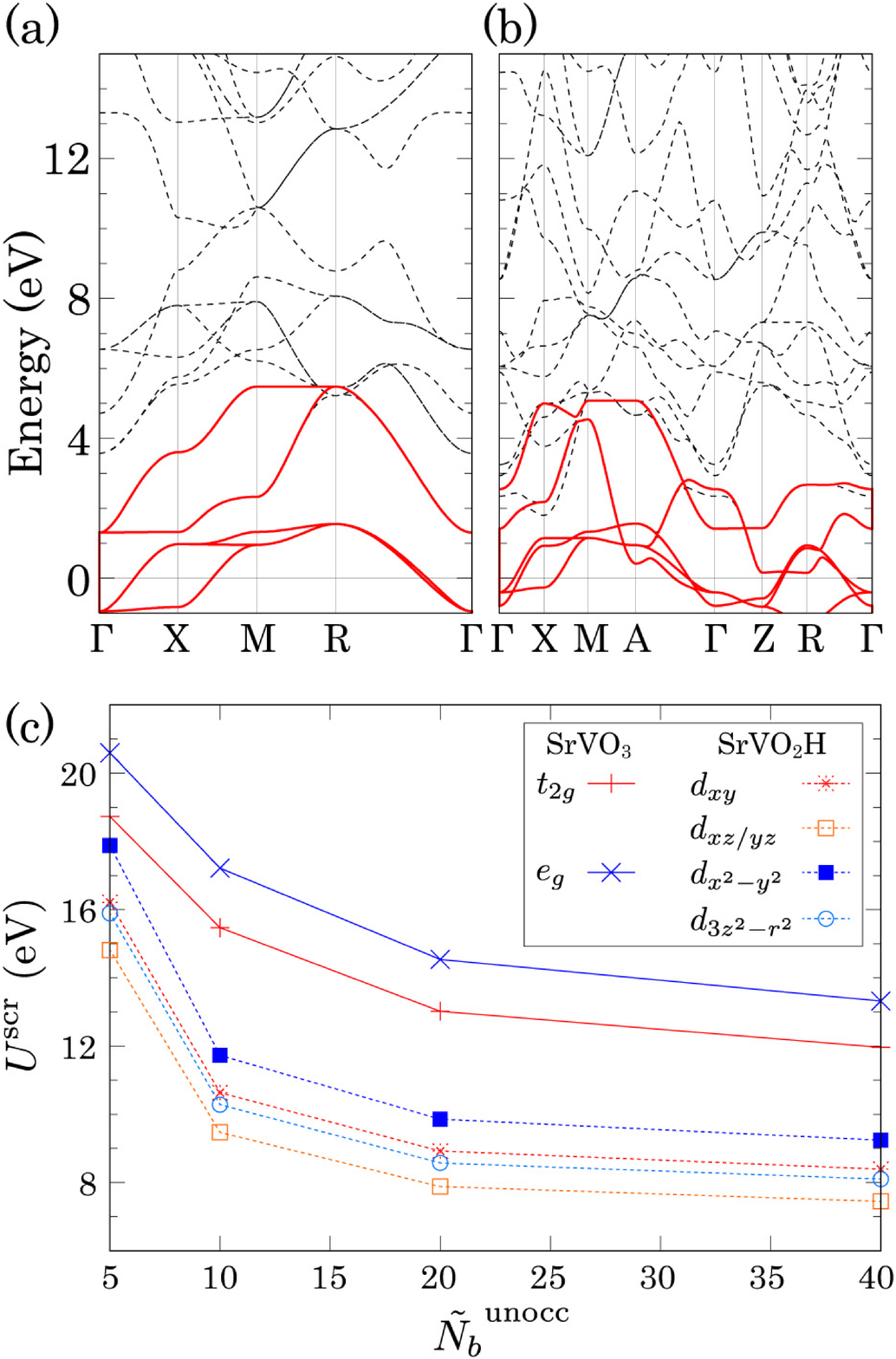}
\caption{(a) High-energy region of the electronic band structure for SrVO$_3$ shown in Fig.~\ref{fig:band1}(c). (b) The same plot for SrVO$_2$H, i.e., Fig.~\ref{fig:band1}(f). (c) $\tilde{N}_b^{\mathrm{unocc}}$-dependence of the screened interaction parameter $U^{\mathrm{scr}}$ in the $dp(s)$ model. The definition of $\tilde{N}_b^{\mathrm{unocc}}$ is given in the main text.}
\label{fig:nbdep}
\end{center}
\end{figure}

From these results, we found that the screened interaction in the $dp(s)$ model is much smaller for SrVO$_2$H than that for SrVO$_3$.
One of the reasons is the extended character of the Wannier functions in SrVO$_2$H as we have seen, which can be inferred from the smaller bare interaction $U^{\mathrm{bare}}_{dps}$ in SrVO$_2$H than $U^{\mathrm{bare}}_{dp}$ in SrVO$_3$.
Another reason is the fact that there are many high-energy bands close to the V-$d$ bands in SrVO$_2$H as shown in Fig.~\ref{fig:nbdep}(b), compared with SrVO$_3$ as shown in Fig.~\ref{fig:nbdep}(a).
Such high-energy bands close to the V-$d$ bands can have a large contribution to the screening process.

To verify this issue, we calculated the screened interaction $U^{\mathrm{scr}}_{dp(s)}$ with changing the number of bands taken into account in cRPA calculations as shown in Fig.~\ref{fig:nbdep}(c).
Here, $\tilde{N}_b^{\mathrm{unocc}}$ roughly corresponds to the number of (partially) unoccupied bands including the V-$3d$ bands.
To be more precise, when one considers the screening processes within the lowest $N_b$ bands, $\tilde{N}_b^{\mathrm{unocc}}$ is defined as $N_b$ subtracted with 20 (i.e., a half of the number of electrons for Sr-$4s4p$, V-$3s3p$, and O-$2s2p$) in SrVO$_3$ and 17 (i.e., a half of the number of electrons for Sr-$4s4p$, V-$3s3p$, O-$2s2p$, and H-$1s$) in SrVO$_2$H.
A sharp drop of $U^{\mathrm{scr}}$ at small $\tilde{N}_b^{\mathrm{unocc}}$ for SrVO$_2$H, as shown in Fig.~\ref{fig:nbdep}(c),
suggests that the strong entanglement of the V-$d$ bands and higher-energy bands in SrVO$_2$H is important for the strong screening effects.

For the $d$ model, the screened Coulomb interaction among $t_{2g}$ orbitals in SrVO$_2$H now becomes stronger than SrVO$_3$, as shown in Table~\ref{table:113_d}.
The difference between the $d$ and $dp(s)$ models should come from the screening effects by the O-$p$ and H-$s$ orbitals:
i.e., these anion orbitals weakly screen the Coulomb interaction among the V-$d$ orbitals in SrVO$_2$H compared with SrVO$_3$.
This is naturally expected by the band dispersion shown in Fig.~\ref{fig:band1}(b)(e), where the anion bands are more separated from the V-$d$ bands in SrVO$_2$H than SrVO$_3$. In fact, the stronger screening effect in SrCrO$_3$ than SrVO$_3$ shown in Tables~\ref{table:113_t2g} and \ref{table:113_d} originates from a smaller energy difference between the O-$p$ and V(Cr)-$d$ bands, as pointed out in Ref.~\onlinecite{Vaugier_crpa}.
Here, the lower on-site energy of Cr-$d$ than that for V-$d$ owing to the increased nuclear charge for Cr is the origin of such a small energy difference in SrCrO$_3$.
It was theoretically pointed out that a similar situation was realized also in cuprates, where the longer the bond distance between apical oxygen and copper is, the stronger the screening effect becomes owing to a smaller $d$-$p$ energy-level deference by stabilization of the copper $d$ orbitals~\cite{SWJ_cup}.
As for SrVO$_2$H, the large energy separation between the anion bands and the V-$d$ bands is likely to come from the fact that the O-$p$ orbitals are more stabilized by the existence of hydrogen atoms, compared with the V-$d$ orbitals. As a matter of fact, the on-site energy difference between the V-$d_{yz}$ and O-$p_z$ orbitals in the $dp(s)$ model for SrVO$_3$ and SrVO$_2$H is 3.31 and 3.97 eV, respectively.
It is also important that the number of the O-$p$ bands is reduced in SrVO$_2$H from SrVO$_3$.
As for the $e_g$ orbitals, it is noteworthy that both the bare and screened Coulomb interaction parameters are weak for the $d_{3z^2-r^2}$ orbitals in SrVO$_2$H as shown in Table~\ref{table:113_d}, because of its extended nature that we have seen in previous sections. 

Finally, we come to the $t_{2g}$ model.
The effective interaction in SrVO$_2$H is again, as in the $dp(s)$ model, smaller than that in SrVO$_3$, likely because of the strong entanglement among the $t_{2g}$ and $e_g$ bands. In other words, the $e_g$ orbitals strongly screen the effective interaction among the $t_{2g}$ orbitals in SrVO$_2$H.
In fact, the difference between $U^{\mathrm{scr}}_{t_{2g}}$ and $U^{\mathrm{scr}}_d$ is negligible in SrVO$_3$ (3.42 and 3.43 eV),
very small in SrCrO$_3$ (2.97 and 3.04 eV) where the $t_{2g}$ and $e_g$ bands are slightly entangled,
and quite large in SrVO$_2$H (3.00 and 3.97 eV for $d_{xy}$, 2.60 and 3.75 eV for $d_{xz/yz}$) where the $t_{2g}$ and $e_g$ bands are strongly entangled.
We note that the bare Coulomb interaction $U^{\mathrm{bare}}_{t_{2g}}$ and $U^{\mathrm{bare}}_d$ are rather close in SrVO$_2$H: 16.04 and 16.06 eV for $d_{xy}$, 15.18 and 15.28 eV for $d_{xz/yz}$, which rules out the possibility that the difference between $U^{\mathrm{scr}}_{t_{2g}}$ and $U^{\mathrm{scr}}_d$ comes from the variation of the Wannier orbitals (e.g., the size of the spread).

We summarize the complicated screening effects in SrVO$_2$H.
The $e_g$ bands strongly entangled with the $t_{2g}$ bands sizably screen the effective interaction in the $t_{2g}$ model.
The large energy separation between the anion bands and the V-$d$ bands weakens the screening effect by the anion orbitals in the $d$ model (and the $t_{2g}$ model).
The strong entanglement of the V-$d$ bands and higher-energy bands yields the strong screening effect in the $dps$ model (and other two effective models).
It is noteworthy that the different $d$-electron numbers between SrVO$_2$H and SrVO$_3$ should play some role for making interaction parameters of these two systems different. As a matter of fact, in Ref.~\onlinecite{nomura_crpa}, the screened interaction ($U$, $U'$, $J$) for the $t_{2g}$ model in SrVO$_3$ evaluated using cRPA was reported to be (3.39, 2.34, 0.47) for the $d^1$ filling and (3.65, 2.59, 0.46) for the $d^2$ filling.
Because the change in $U$ and $U'$ is roughly 0.25 eV here, we can expect that the change in the band structure we have discussed above is still crucial for understanding the peculiar effective interaction in SrVO$_2$H compared with SrVO$_3$.

Because of the sizable difference in effective interaction parameters among the $t_{2g}$ and $d$ models, it is a non-trivial issue which effective model one should adopt for analyzing the electronic structure of SrVO$_2$H. The lowered energy of the $d_{3z^2-r^2}$ bands, which come close to the Fermi energy as shown in Fig.~\ref{fig:band1}(e), might have some relevance to this issue.
We also note that one possible origin of the large difference in the interaction parameters is the difficulty in evaluating $\chi_0^r$ when the band entanglement takes place.
For treating the band entanglement, while we used the method shown in Ref.~\onlinecite{cRPA_entangle} as implemented in the RESPACK code, there is another choice such as the one shown in Ref.~\onlinecite{Miyake_entangle}. When the metallic screening is not fully removed by using the former method, the interaction parameters will become small (i.e., overscreened), which can be related to the case of the $t_{2g}$ model in SrVO$_2$H.
These are important future problems.

\subsubsection{Off-site direct Coulomb interaction}

We found that the off-site direct Coulomb interaction parameters exhibit a similar tendency to the on-site parameters for each model.
For example, the screened interaction parameters of the nearest-neighbor off-site direct Coulomb interaction along the $z$ direction for the $d$ model are (0.58, 0.74) eV in SrVO$_3$ and (0.72, 0.83) eV in SrVO$_2$H, where the orbital-diagonal components of the interaction parameters for the $d_{xy}$ and $d_{xz/yz}$ orbitals are shown.
Similarly to the on-site terms, the effective off-site interaction in SrVO$_2$H is stronger than that in SrVO$_3$ for the $d$ model.
On the other hand, the corresponding off-site interaction parameters become (0.58, 0.74) eV in SrVO$_3$ and (0.27, 0.33) eV in SrVO$_2$H, for the $t_{2g}$ model, where the on-site screened interaction in SrVO$_2$H is also weaker than that in SrVO$_3$.
The above results might come from the fact that it is unchanged which electron excitations tend to play a major role in the screening process, irrespective of whether it is on-site or off-site interactions.

\subsubsection{Exchange interaction}

Unlike other parameters, the exchange interaction $J$ is less sensitive to the existence of hydrogen atoms.
For example, $J^{\mathrm{scr}}_{t_{2g}}$ is 0.48 eV for SrVO$_3$, while it is 0.46 between the $d_{xy}$ and $d_{xz/yz}$ orbitals, and 0.42 between the $d_{xz}$ and $d_{yz}$ orbitals, for SrVO$_2$H.
An unusual aspect of SrVO$_2$H is a relatively large off-site screened exchange interaction between $d_{3z^2-r^2}$ and H-$s$ in the $dps$ model, 0.16 eV, while all the other off-site screened exchange interaction is less than 0.1 eV in SrVO$_3$ and SrVO$_2$H.
This might be due to the large overlap of these two orbitals and the shortened lattice constant along the $z$ direction.

\subsection{Sr$_2$VO$_4$ and Sr$_2$VO$_3$H ($n=1$)\label{sec:results214}}

\begin{figure}
\begin{center}
\includegraphics[width=8.5 cm]{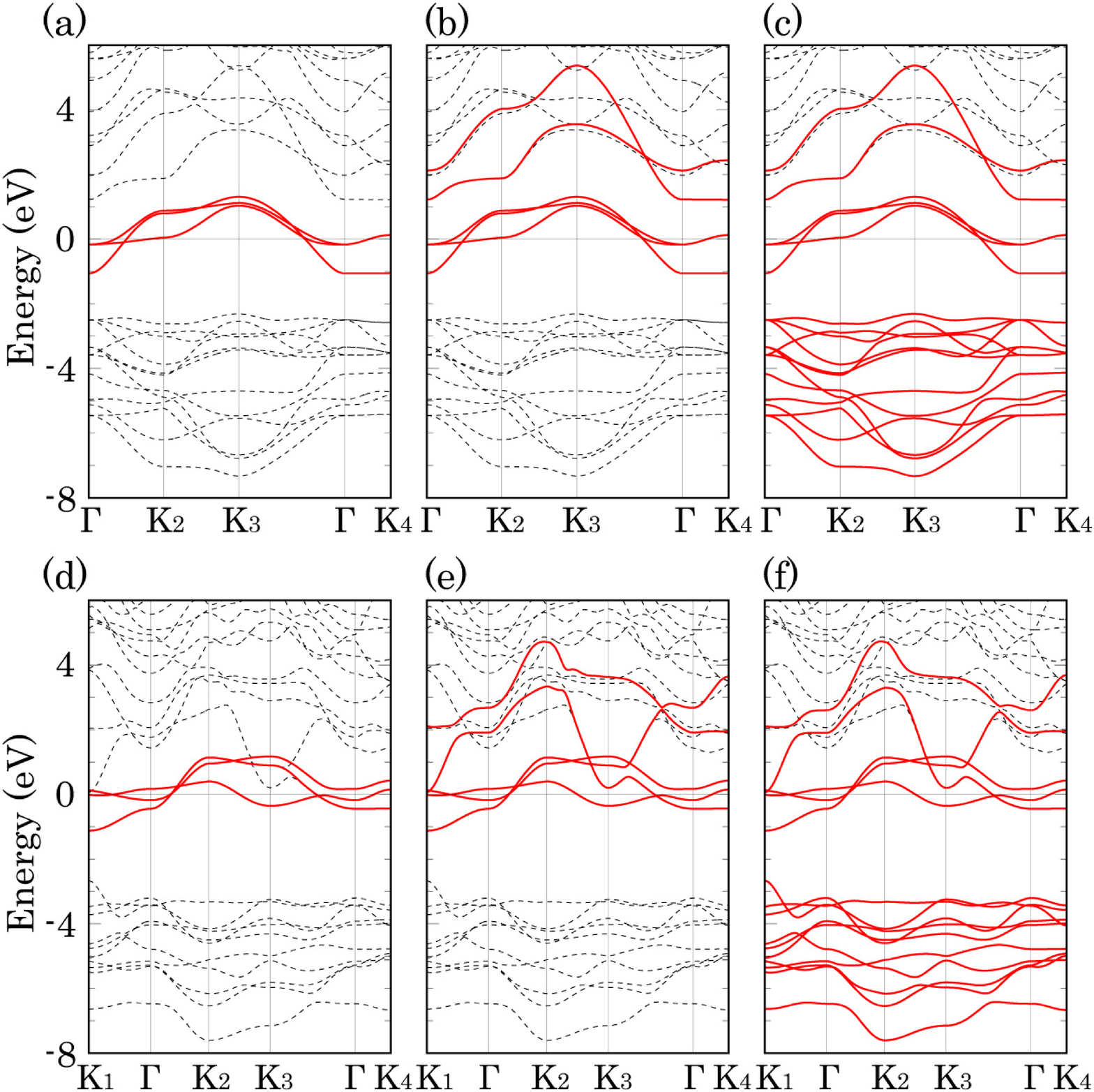}
\caption{Calculated electronic band structure of (a)--(c) Sr$_2$VO$_4$ and (d)--(f) Sr$_2$VO$_3$H. First-principles band structure is shown with black broken lines and the band dispersion calculated with the tight-binding model consisting of the Wannier functions is shown with red solid lines. Corresponding effective models are $t_{2g}$ for panels (a) and (d), $d$ for panels (b) and (e), $dp$ for panel (c), and $dps$ for panel (f). To compare the band structures of Sr$_2$VO$_4$ and Sr$_2$VO$_3$H, common special $k$-points were taken as follows: K$_1$ = ($\pi/a$, 0, 0), K$_2$ = (0, $\pi/b$, 0), K$_3$ = ($\pi/a$, $\pi/b$, 0), and K$_4$= (0, 0, $2\pi/c$) in the cartesian coordinate, where $a$, $b$ ($=a$ for Sr$_2$VO$_4$) and $c$ are the lattice constants shown in Fig.~\ref{fig:crys}(c)--(d).}
\label{fig:band2}
\end{center}
\end{figure}

For Sr$_2$VO$_4$ and Sr$_2$VO$_3$H ($n=1$), we considered the $d_{y^2-z^2}$ and $d_{3x^2-r^2}$ orbitals as the $e_g$ orbitals, instead of the $d_{x^2-y^2}$ and $d_{3z^2-r^2}$ orbitals.
This is because hydrogen atoms make the $x$ axis quite inequivalent from other axes in Sr$_2$VO$_3$H, as shown in Fig.~\ref{fig:crys}(d).

\begin{table}
\begin{center}
\begin{tabular}{c c c c c c c}
\hline \hline
 & & $t_x$ & $t_y$ &  $\Delta$ & $U^{\mathrm{scr}}_{t_{2g}}$ & $U^{\mathrm{bare}}_{t_{2g}}$  \\
\hline
Sr$_2$VO$_4$ & $d_{xy}$ & $-0.27$ & $-0.27$ & - & 3.46 & 15.91\\
& & & & & 2.77$^*$ [\onlinecite{214_crpa1,214_crpa2}] &  \\
& & & & & 3.1$^{**}$ [\onlinecite{casula_crpa}] &  \\
& $d_{yz}$ & $-0.04$ & $-0.24$ & $-0.02$ & 3.26 & 15.18 \\
& & & & & 2.58$^*$ [\onlinecite{214_crpa1,214_crpa2}] &  \\
& & & & & 3.1$^{**}$ [\onlinecite{casula_crpa}] &  \\
 \hline
 Sr$_2$VO$_3$H & $d_{xy}$ & 0.10 & $-0.44$ & - & 2.51 & 14.93 \\
 & $d_{yz}$ & $-0.06$ & $-0.25$ & 0.39 & 2.84 & 15.25 \\
 & $d_{xz}$ & 0.14 & 0.03 & $-0.05$ & 2.46 & 14.26 \\
\hline \hline
\end{tabular}
\end{center}
\caption{\label{table:214_t2g} Hopping and interaction parameters (in eV) for the $t_{2g}$ model. The $d_{xz}$ orbital in Sr$_2$VO$_4$ is omitted in this table since it is equivalent to the $d_{yz}$ orbital.
$^*$: constrained LDA combined with the $GW$ method.
 $^{**}$: averaged over the orbitals.}
\end{table}

\begin{table}
\begin{center}
\begin{tabular}{c c c c c c c}
\hline \hline
 & & $t_x$ & $t_y$ & $\Delta$ & $U^{\mathrm{scr}}_d$ & $U^{\mathrm{bare}}_d$ \\
\hline
Sr$_2$VO$_4$ & $d_{xy}$ & $-0.27$ & $-0.27$ & - & 3.48 & 15.95 \\
& $d_{yz}$ & $-0.04$ & $-0.24$ & $-0.02$ & 3.27 & 15.19 \\
 & $d_{y^2-z^2}$ & $-0.01$ & $-0.47$ & 2.61 & 3.33 & 15.13 \\
 & $d_{3x^2-r^2}$ & $-0.65$ & $-0.18$ & 2.73 & 3.49 & 15.98 \\
 \hline
 Sr$_2$VO$_3$H & $d_{xy}$ & 0.09 & $-0.45$ & - & 3.60 & 15.20 \\
 & $d_{yz}$ & $-0.06$ & $-0.25$ & 0.38 & 3.70 & 15.49 \\
  & $d_{xz}$ & 0.14 & 0.03 & $-0.07$ & 3.36 & 14.25 \\
 & $d_{y^2-z^2}$ & 0.00 & $-0.43$ & 2.91 & 3.86 & 16.04  \\
 & $d_{3x^2-r^2}$ & 0.88 & $-0.09$ & 1.90 & 3.14 & 13.56 \\
\hline \hline
\end{tabular}
\end{center}
\caption{\label{table:214_d} Hopping and interaction parameters (in eV) for the $d$ model. The $d_{xz}$ orbital in Sr$_2$VO$_4$ is omitted in this table since it is equivalent to the $d_{yz}$ orbital.}
\end{table}

Because the situation is basically similar between $n=1$ and $n=\infty$, we just briefly show our calculation results. 
Figure~\ref{fig:band2} presents the calculated band structure of Sr$_2$VO$_4$ and Sr$_2$VO$_3$H.
We can find some features similar to the $n=\infty$ case: strong entanglement of the $t_{2g}$ and $e_g$ bands, and a large energy separation between the V-$d$ and anion bands.
As a result, the screening interaction of Sr$_2$VO$_3$H is weaker in the $t_{2g}$ model but stronger in the $d$ model, compared with Sr$_2$VO$_4$.
An extended nature of the $d_{3x^2-r^2}$ orbital, which forms a chemical bond with the H-$s$ orbital, is also realized in Sr$_2$VO$_3$H.
In fact, Table~\ref{table:214_d} shows that both $U^{\mathrm{scr}}_d$ and $U^{\mathrm{bare}}_d$ exhibit small values for the the $d_{3x^2-r^2}$ orbital in Sr$_2$VO$_3$H.

One important difference between $n=1$ and $n=\infty$ is the dimensionality of the electronic structure.
As shown in Table~\ref{table:214_t2g}, one can see that the hopping parameter along the $x$ direction, $t_x$, is actually suppressed in SrVO$_3$H, where the hydrogen atom breaks a chemical bond between the V-$t_{2g}$ orbitals.
In addition, the layered structure cannot yield a large $t_z$ both for Sr$_2$VO$_4$ and Sr$_2$VO$_3$H.
Therefore, the $t_{2g}$ orbitals in Sr$_2$VO$_3$H have a peculiar dimensionality.
As for the $d_{xy}$ orbital, its hopping parameters are quite similar to those for the $d_{yz}$ orbital in SrVO$_2$H shown in Table~\ref{table:113_t2g}.
In other words, the $d_{xy}$ orbital in Sr$_2$VO$_3$H has quasi-one-dimensional electronic structure with an enhanced hopping amplitude along the $y$ direction.
As for the $d_{yz}$ orbital, it is less affected by the existence of hydrogen atoms, except for the on-site energy difference with the $d_{xy}$ orbital, $\Delta$, as shown in Table~\ref{table:214_t2g}.
The $d_{xz}$ orbital in Sr$_2$VO$_3$H has a small hopping amplitude for all the directions, which is unique for $n=1$.

\section{Conclusion\label{sec:sum}}

We have derived several kinds of low-energy effective models for vanadium oxyhydrides Sr$_{n+1}$V$_n$O$_{2n+1}$H$_n$ ($n=1,\infty$) and some oxides: SrVO$_3$, SrCrO$_3$, and Sr$_2$VO$_4$, using cRPA.
We have found that, in SrVO$_2$H, (1) the $e_g$ bands strongly entangled with the $t_{2g}$ bands sizably screen the effective interaction in the $t_{2g}$ model,
(2) the large energy separation between the anion bands and the V-$d$ bands weakens the screening effect by the anion orbitals in the $d$ model (and the $t_{2g}$ model), and
(3) the strong entanglement of the V-$d$ bands and higher-energy bands yields the strong screening effect in the $dps$ model (and other two effective models).
A similar tendency can be seen also in Sr$_2$VO$_3$H.
Investigation of possible unique correlation effects in vanadium oxyhydrides based on the low-energy effective models derived in the present 
study is an open and interesting future study.

\acknowledgments
Some calculations were performed using large-scale computer systems in the supercomputer center of the Institute for Solid State Physics, the University of Tokyo, and those of the Cybermedia Center, Osaka University.
This study was supported by JSPS KAKENHI (Grant Nos.~JP16H04338, JP17H05481, JP18H01860, and JP18K13470), Japan.

\appendix
\section{Hopping parameters}

The orbital index runs as ($d_{xy}$, $d_{yz}$, $d_{xz}$) for the $t_{2g}$ model,
and ($d_{xy}$, $d_{yz}$, $d_{xz}$, $d_{x^2-y^2}$, $d_{3z^2-r^2}$) for the $d$ model, unless noted.
The hopping parameters are defined for the (non-interacting) tight-binding Hamiltonian:
\begin{equation}
\mathcal{H}_0 = \sum_{ij} t_{ij} (\mathbf{R}) \hat{c}_i^{\dag} \hat{c}_j,
\end{equation}
where $i$ and $j$ are orbital indices. For clarity, we represent the lattice vector $\mathbf{R}$ with the Cartesian coordinate defined in Fig.~\ref{fig:crys}.
For the $t_{2g}$ and $d$ models, we denote the hopping matrix as $t^{t_{2g}} (R_x, R_y, R_z)$ and $t^{d} (R_x, R_y, R_z)$, respectively.

Some equivalent parameters are omitted here. For example, in SrVO$_3$, $t^{t_{2g}} (a,0,0)$, $t^{t_{2g}} (0,a,0)$, and $t^{t_{2g}} (0,0,a)$ are equivalent if the orbital indices are appropriately exchanged, and thus we only show one of them.
For all the compounds investigated in this study, the $d_{xy}$ onsite energy is set to zero.

\subsection{SrVO$_3$}

\begin{equation}
t^{t_{2g}} (0,0,0) = \left(
\begin{array}{ccc}
0\ &\ \ 0\ &\ \ 0 \\
0\ &\ \ 0\ &\ \ 0 \\
0\ &\ \ 0\ &\ \ 0
\end{array}
\right),
\end{equation}

\begin{equation}
t^{t_{2g}} (a,0,0) = \left(
\begin{array}{ccc}
-0.263\ &\ 0\ &\ 0 \\
0\ &\ -0.027\ &\ 0 \\
0\ &\ 0\ &\ -0.263
\end{array}
\right),
\end{equation}

\begin{equation}
t^{t_{2g}} (a,a,0) = \left(
\begin{array}{ccc}
-0.084\ &\ 0\ &\ 0 \\
0\ &\ 0.006\ &\ 0.009 \\
0\ &\ 0.009\ &\ 0.006
\end{array}
\right),
\end{equation}

\begin{equation}
t^{d} (0,0,0) = \left(
\begin{array}{ccccc}
0\ &\ \ 0\ &\ \ 0\ &\ \ 0\ &\ \ 0 \\
0\ &\ \ 0\ &\ \ 0\ &\ \ 0\ &\ \ 0 \\
0\ &\ \ 0\ &\ \ 0\ &\ \ 0\ &\ \ 0 \\
0\ &\ \ 0\ &\ \ 0\ &\ \ 2.765\ &\ \ 0 \\
0\ &\ \ 0\ &\ \ 0\ &\ \ 0\ &\ \ 2.765
\end{array}
\right),
\end{equation}

\begin{equation}
t^{d} (a,0,0) = \left(
\begin{array}{ccccc}
-0.262\ &\ 0\ &\ 0\ &\ 0\ &\ 0 \\
0\ &\ -0.025\ &\ 0\ &\ 0\ &\ 0 \\
0\ &\ 0\ &\ -0.262\ &\ 0\ &\ 0 \\
0\ &\ 0\ &\ 0\ &\ -0.505\ &\ 0.293 \\
0\ &\ 0\ &\ 0\ &\ 0.293\ &\ -0.167
\end{array}
\right),
\end{equation}

\begin{equation}
t^{d} (0,0,a) = \left(
\begin{array}{ccccc}
-0.025\ &\ 0\ &\ 0\ &\ 0\ &\ 0 \\
0\ &\ -0.262\ &\ 0\ &\ 0\ &\ 0 \\
0\ &\ 0\ &\ -0.262\ &\ 0\ &\ 0 \\
0\ &\ 0\ &\ 0\ &\ 0.003\ &\ 0 \\
0\ &\ 0\ &\ 0\ &\ 0\ &\ -0.674
\end{array}
\right),
\end{equation}

\begin{equation}
t^{d} (2a,0,0) = \left(
\begin{array}{ccccc}
0.005\ &\ 0\ &\ 0\ &\ 0\ &\ 0 \\
0\ &\ 0.001\ &\ 0\ &\ 0\ &\ 0 \\
0\ &\ 0\ &\ 0.005\ &\ 0\ &\ 0 \\
0\ &\ 0\ &\ 0\ &\ -0.039\ &\ 0.023 \\
0\ &\ 0\ &\ 0\ &\ 0.023\ &\ -0.013
\end{array}
\right),
\end{equation}

\begin{equation}
t^{d} (0,0,2a) = \left(
\begin{array}{ccccc}
0.001\ &\ 0\ &\ 0\ &\ 0\ &\ 0 \\
0\ &\ 0.005\ &\ 0\ &\ 0\ &\ 0 \\
0\ &\ 0\ &\ 0.005\ &\ 0\ &\ 0 \\
0\ &\ 0\ &\ 0\ &\ 0.000\ &\ 0 \\
0\ &\ 0\ &\ 0\ &\ 0\ &\ -0.052
\end{array}
\right),
\end{equation}

\begin{equation}
t^{d} (a,a,0) = \left(
\begin{array}{ccccc}
-0.083\ &\ 0\ &\ 0\ &\ 0\ &\ -0.031 \\
0\ &\ 0.006\ &\ 0.010\ &\ 0\ &\ 0 \\
0\ &\ 0.010\ &\ 0.006\ &\ 0\ &\ 0 \\
0\ &\ 0\ &\ 0\ &\ 0.041\ &\ 0 \\
-0.031\ &\ 0\ &\ 0\ &\ 0\ &\ -0.017
\end{array}
\right),
\end{equation}

\begin{equation}
t^{d} (a,0,a) = \left(
\begin{array}{ccccc}
0.006\ &\ 0.010\ &\ 0\ &\ 0\ &\ 0 \\
0.010\ &\ 0.006\ &\ 0\ &\ 0\ &\ 0 \\
0\ &\ 0\ &\ -0.083\ &\ 0.027\ &\ 0.015 \\
0\ &\ 0\ &\ 0.027\ &\ -0.002\ &\ -0.025 \\
0\ &\ 0\ &\ 0.015\ &\ -0.025\ &\ 0.027
\end{array}
\right).
\end{equation}

\subsection{SrVO$_2$H}

\begin{equation}
t^{t_{2g}} (0,0,0) = \left(
\begin{array}{ccc}
0\ &\ \ 0\ &\ \ 0 \\
0\ &\ \ -0.454\ &\ \ 0 \\
0\ &\ \ 0\ &\ \ -0.454
\end{array}
\right),
\end{equation}

\begin{equation}
t^{t_{2g}} (a,0,0) = \left(
\begin{array}{ccc}
-0.251\ &\ 0\ &\ 0 \\
0\ &\ 0.012\ &\ 0 \\
0\ &\ 0\ &\ -0.423
\end{array}
\right),
\end{equation}

\begin{equation}
t^{t_{2g}} (0,0,c) = \left(
\begin{array}{ccc}
-0.040\ &\ 0\ &\ 0 \\
0\ &\ 0.097\ &\ 0 \\
0\ &\ 0\ &\ 0.097
\end{array}
\right),
\end{equation}

\begin{equation}
t^{t_{2g}} (0,0,2c) = \left(
\begin{array}{ccc}
-0.001\ &\ 0\ &\ 0 \\
0\ &\ 0.017\ &\ 0 \\
0\ &\ 0\ &\ 0.017
\end{array}
\right),
\end{equation}

\begin{equation}
t^{t_{2g}} (a,a,0) = \left(
\begin{array}{ccc}
-0.068\ &\ 0\ &\ 0 \\
0\ &\ 0.013\ &\ 0.027 \\
0\ &\ 0.027\ &\ 0.013
\end{array}
\right),
\end{equation}

\begin{equation}
t^{t_{2g}} (a,0,c) = \left(
\begin{array}{ccc}
0.002\ &\ -0.009\ &\ 0 \\
-0.009\ &\ -0.010\ &\ 0 \\
0\ &\ 0\ &\ 0.027
\end{array}
\right),
\end{equation}

\begin{equation}
t^{d} (0,0,0) = \left(
\begin{array}{ccccc}
0\ &\ 0\ &\ 0\ &\ 0\ &\ 0 \\
0\ &\ -0.443\ &\ 0\ &\ 0\ &\ 0 \\
0\ &\ 0\ &\ -0.443\ &\ 0\ &\ 0 \\
0\ &\ 0\ &\ 0\ &\ 2.489\ &\ 0 \\
0\ &\ 0\ &\ 0\ &\ 0\ &\ 1.523
\end{array}
\right),
\end{equation}

\begin{equation}
t^{d} (a,0,0) = \left(
\begin{array}{ccccc}
-0.251\ &\ 0\ &\ 0\ &\ 0\ &\ 0 \\
0\ &\ 0.008\ &\ 0\ &\ 0\ &\ 0 \\
0\ &\ 0\ &\ -0.424\ &\ 0\ &\ 0 \\
0\ &\ 0\ &\ 0\ &\ -0.438\ &\ 0.183 \\
0\ &\ 0\ &\ 0\ &\ 0.183\ &\ -0.094
\end{array}
\right),
\end{equation}

\begin{equation}
t^{d} (0,0,c) = \left(
\begin{array}{ccccc}
-0.040\ &\ 0\ &\ 0\ &\ 0\ &\ 0 \\
0\ &\ 0.096\ &\ 0\ &\ 0\ &\ 0 \\
0\ &\ 0\ &\ 0.096\ &\ 0\ &\ 0 \\
0\ &\ 0\ &\ 0\ &\ 0.012\ &\ 0 \\
0\ &\ 0\ &\ 0\ &\ 0\ &\ 0.876
\end{array}
\right),
\end{equation}

\begin{equation}
t^{d} (2a,0,0) = \left(
\begin{array}{ccccc}
0.009\ &\ 0\ &\ 0\ &\ 0\ &\ 0 \\
0\ &\ 0.001\ &\ 0\ &\ 0\ &\ 0 \\
0\ &\ 0\ &\ -0.017\ &\ 0\ &\ 0 \\
0\ &\ 0\ &\ 0\ &\ -0.035\ &\ 0.020 \\
0\ &\ 0\ &\ 0\ &\ 0.020\ &\ -0.004
\end{array}
\right),
\end{equation}

\begin{equation}
t^{d} (0,0,2c) = \left(
\begin{array}{ccccc}
-0.001\ &\ 0\ &\ 0\ &\ 0\ &\ 0 \\
0\ &\ 0.010\ &\ 0\ &\ 0\ &\ 0 \\
0\ &\ 0\ &\ 0.010\ &\ 0\ &\ 0 \\
0\ &\ 0\ &\ 0\ &\ -0.001\ &\ 0 \\
0\ &\ 0\ &\ 0\ &\ 0\ &\ -0.059
\end{array}
\right),
\end{equation}

\begin{equation}
t^{d} (a,a,0) = \left(
\begin{array}{ccccc}
-0.068\ &\ 0\ &\ 0\ &\ 0\ &\ -0.007 \\
0\ &\ 0.013\ &\ 0.022\ &\ 0\ &\ 0 \\
0\ &\ 0.022\ &\ 0.013\ &\ 0\ &\ 0 \\
0\ &\ 0\ &\ 0\ &\ 0.058\ &\ 0 \\
-0.007\ &\ 0\ &\ 0\ &\ 0\ &\ -0.041
\end{array}
\right),
\end{equation}

\begin{equation}
t^{d} (a,0,c) = \left(
\begin{array}{ccccc}
0.002\ &\ -0.010\ &\ 0\ &\ 0\ &\ 0 \\
-0.010\ &\ -0.010\ &\ 0\ &\ 0\ &\ 0 \\
0\ &\ 0\ &\ 0.027\ &\ -0.014\ &\ -0.084 \\
0\ &\ 0\ &\ -0.014\ &\ 0.001\ &\ 0.054 \\
0\ &\ 0\ &\ -0.084\ &\ 0.054\ &\ -0.064
\end{array}
\right),
\end{equation}

\begin{equation}
t^{d} (0,0,3c) = \left(
\begin{array}{ccccc}
0.000\ &\ 0\ &\ 0\ &\ 0\ &\ 0 \\
0\ &\ 0.000\ &\ 0\ &\ 0\ &\ 0 \\
0\ &\ 0\ &\ 0.000\ &\ 0\ &\ 0 \\
0\ &\ 0\ &\ 0\ &\ 0.001\ &\ 0 \\
0\ &\ 0\ &\ 0\ &\ 0\ &\ 0.011
\end{array}
\right),
\end{equation}

\begin{equation}
t^{d} (a,0,2c) = \left(
\begin{array}{ccccc}
0.000\ &\ -0.001\ &\ 0\ &\ 0\ &\ 0 \\
-0.001\ &\ -0.001\ &\ 0\ &\ 0\ &\ 0 \\
0\ &\ 0\ &\ 0.001\ &\ -0.001\ &\ 0.015 \\
0\ &\ 0\ &\ -0.001\ &\ 0.001\ &\ 0.000 \\
0\ &\ 0\ &\ 0.015\ &\ 0.000 &\ -0.030
\end{array}
\right),
\end{equation}

\begin{equation}
t^{d} (a,a,c) = \left(
\begin{array}{ccccc}
-0.006\ &\ -0.004\ &\ -0.004\ &\ 0\ &\ 0.001 \\
-0.004\ &\ -0.010\ &\ 0.006\ &\ -0.009\ &\ 0.015 \\
-0.004\ &\ 0.006\ &\ -0.010\ &\ 0.009\ &\ 0.015 \\
0\ &\ -0.009\ &\ 0.009\ &\ -0.011\ &\ 0 \\
0.001\ &\ 0.015\ &\ 0.015\ &\ 0 &\ -0.027
\end{array}
\right).
\end{equation}

\subsection{SrCrO$_3$}

\begin{equation}
t^{t_{2g}} (0,0,0) = \left(
\begin{array}{ccc}
0\ &\ \ 0\ &\ \ 0 \\
0\ &\ \ 0\ &\ \ 0 \\
0\ &\ \ 0\ &\ \ 0
\end{array}
\right),
\end{equation}

\begin{equation}
t^{t_{2g}} (a,0,0) = \left(
\begin{array}{ccc}
-0.236\ &\ 0\ &\ 0 \\
0\ &\ -0.022\ &\ 0 \\
0\ &\ 0\ &\ -0.236
\end{array}
\right),
\end{equation}

\begin{equation}
t^{t_{2g}} (a,a,0) = \left(
\begin{array}{ccc}
-0.087\ &\ 0\ &\ 0 \\
0\ &\ 0.008\ &\ 0.010 \\
0\ &\ 0.010\ &\ 0.008
\end{array}
\right),
\end{equation}

\begin{equation}
t^{d} (0,0,0) = \left(
\begin{array}{ccccc}
0\ &\ \ 0\ &\ \ 0\ &\ \ 0\ &\ \ 0 \\
0\ &\ \ 0\ &\ \ 0\ &\ \ 0\ &\ \ 0 \\
0\ &\ \ 0\ &\ \ 0\ &\ \ 0\ &\ \ 0 \\
0\ &\ \ 0\ &\ \ 0\ &\ \ 2.522\ &\ \ 0 \\
0\ &\ \ 0\ &\ \ 0\ &\ \ 0\ &\ \ 2.522
\end{array}
\right),
\end{equation}

\begin{equation}
t^{d} (a,0,0) = \left(
\begin{array}{ccccc}
-0.236\ &\ 0\ &\ 0\ &\ 0\ &\ 0 \\
0\ &\ -0.022\ &\ 0\ &\ 0\ &\ 0 \\
0\ &\ 0\ &\ -0.236\ &\ 0\ &\ 0 \\
0\ &\ 0\ &\ 0\ &\ -0.509\ &\ 0.295 \\
0\ &\ 0\ &\ 0\ &\ 0.295\ &\ -0.168
\end{array}
\right),
\end{equation}

\begin{equation}
t^{d} (0,0,a) = \left(
\begin{array}{ccccc}
-0.022\ &\ 0\ &\ 0\ &\ 0\ &\ 0 \\
0\ &\ -0.236\ &\ 0\ &\ 0\ &\ 0 \\
0\ &\ 0\ &\ -0.236\ &\ 0\ &\ 0 \\
0\ &\ 0\ &\ 0\ &\ 0.002\ &\ 0 \\
0\ &\ 0\ &\ 0\ &\ 0\ &\ -0.679
\end{array}
\right),
\end{equation}

\begin{equation}
t^{d} (2a,0,0) = \left(
\begin{array}{ccccc}
0.004\ &\ 0\ &\ 0\ &\ 0\ &\ 0 \\
0\ &\ 0.000\ &\ 0\ &\ 0\ &\ 0 \\
0\ &\ 0\ &\ 0.004\ &\ 0\ &\ 0 \\
0\ &\ 0\ &\ 0\ &\ -0.045\ &\ 0.026 \\
0\ &\ 0\ &\ 0\ &\ 0.026\ &\ -0.015      
\end{array}
\right),
\end{equation}

\begin{equation}
t^{d} (0,0,2a) = \left(
\begin{array}{ccccc}
0.000\ &\ 0\ &\ 0\ &\ 0\ &\ 0 \\
0\ &\ 0.004\ &\ 0\ &\ 0\ &\ 0 \\
0\ &\ 0\ &\ 0.004\ &\ 0\ &\ 0 \\
0\ &\ 0\ &\ 0\ &\ 0.000\ &\ 0 \\
0\ &\ 0\ &\ 0\ &\ 0\ &\ -0.060      
\end{array}
\right),
\end{equation}

\begin{equation}
t^{d} (a,a,0) = \left(
\begin{array}{ccccc}
-0.087\ &\ 0\ &\ 0\ &\ 0\ &\ -0.032 \\
0\ &\ 0.008\ &\ 0.010\ &\ 0\ &\ 0 \\
0\ &\ 0.010\ &\ 0.008\ &\ 0\ &\ 0 \\
0\ &\ 0\ &\ 0\ &\ 0.044\ &\ 0 \\
-0.032\ &\ 0\ &\ 0\ &\ 0\ &\ -0.017
\end{array}
\right),
\end{equation}

\begin{equation}
t^{d} (a,0,a) = \left(
\begin{array}{ccccc}
0.008\ &\ 0.010\ &\ 0\ &\ 0\ &\ 0 \\
0.010\ &\ 0.008\ &\ 0\ &\ 0\ &\ 0 \\
0\ &\ 0\ &\ -0.087\ &\ 0.028\ &\ 0.016 \\
0\ &\ 0\ &\ 0.028\ &\ -0.001\ &\ -0.026 \\
0\ &\ 0\ &\ 0.016\ &\ -0.026\ &\ 0.029
\end{array}
\right).
\end{equation}

\subsection{Sr$_2$VO$_4$}

\begin{equation}
t^{t_{2g}} (0,0,0) = \left(
\begin{array}{ccc}
0\ &\ \ 0\ &\ \ 0 \\
0\ &\ \ -0.017\ &\ \ 0 \\
0\ &\ \ 0\ &\ \ -0.017
\end{array}
\right),
\end{equation}

\begin{equation} 
t^{t_{2g}} (a,0,0) = \left(
\begin{array}{ccc}
-0.272\ &\ 0\ &\ 0 \\
0\ &\ -0.045\ &\ 0 \\
0\ &\ 0\ &\ -0.240
\end{array}
\right),
\end{equation}

\begin{equation} 
t^{t_{2g}} (2a,0,0) = \left(
\begin{array}{ccc}
0.006\ &\ 0\ &\ 0 \\
0\ &\ 0.001\ &\ 0 \\
0\ &\ 0\ &\ 0.023
\end{array}
\right),
\end{equation}

\begin{equation} 
t^{t_{2g}} (a,a,0) = \left(
\begin{array}{ccc}
-0.082\ &\ 0\ &\ 0 \\
0\ &\ 0.006\ &\ 0.000 \\
0\ &\ 0.000\ &\ 0.006
\end{array}
\right),
\end{equation}

\begin{equation} 
t^{t_{2g}} (\frac{a}{2},\frac{a}{2},\frac{c}{2}) = \left(
\begin{array}{ccc}
0.001\ &\ 0.003\ &\ 0.003 \\
0.003\ &\ -0.015\ &\ -0.011 \\
0.003\ &\ -0.011\ &\ -0.015
\end{array}
\right).
\end{equation}

For the $d$ model in Sr$_2$VO$_4$, the orbital index runs as ($d_{xy}$, $d_{yz}$, $d_{xz}$, $d_{y^2-z^2}$, $d_{3x^2-r^2}$),
in order to compare its effective interaction parameters with those for Sr$_2$VO$_3$H.

\begin{equation}
t^{d} (0,0,0) = \left(
\begin{array}{ccccc}
0\ &\ \ 0\ &\ \ 0\ &\ \ 0\ &\ \ 0 \\
0\ &\ \ -0.020\ &\ \ 0\ &\ \ 0\ &\ \ 0 \\
0\ &\ \ 0\ &\ \ -0.020\ &\ \ 0\ &\ \ 0 \\
0\ &\ \ 0\ &\ \ 0\ &\ \ 2.607\ &\ \ -0.109 \\
0\ &\ \ 0\ &\ \ 0\ &\ \ -0.109\ &\ \ 2.733
\end{array}
\right),
\end{equation}

\begin{equation} 
t^{d} (a,0,0) = \left(
\begin{array}{ccccc}
-0.272\ &\ 0\ &\ 0\ &\ 0\ &\ 0 \\
0\ &\ -0.044\ &\ 0\ &\ 0\ &\ 0 \\
0\ &\ 0\ &\ -0.241\ &\ 0\ &\ 0 \\
0\ &\ 0\ &\ 0\ &\ -0.007\ &\ 0.017 \\
0\ &\ 0\ &\ 0\ &\ 0.017\ &\ -0.650
\end{array}
\right),
\end{equation}

\begin{equation} 
t^{d} (0,a,0) = \left(
\begin{array}{ccccc}
-0.272\ &\ 0\ &\ 0\ &\ 0\ &\ 0 \\
0\ &\ -0.241\ &\ 0\ &\ 0\ &\ 0 \\
0\ &\ 0\ &\ -0.044\ &\ 0\ &\ 0 \\
0\ &\ 0\ &\ 0\ &\ -0.475\ &\ 0.287 \\
0\ &\ 0\ &\ 0\ &\ 0.287\ &\ -0.183
\end{array}
\right),
\end{equation}

\begin{equation} 
t^{d} (2a,0,0) = \left(
\begin{array}{ccccc}
0.005\ &\ 0\ &\ 0\ &\ 0\ &\ 0 \\
0\ &\ 0.001\ &\ 0\ &\ 0\ &\ 0 \\
0\ &\ 0\ &\ 0.022\ &\ 0\ &\ 0 \\
0\ &\ 0\ &\ 0\ &\ 0.000\ &\ 0.002 \\
0\ &\ 0\ &\ 0\ &\ 0.002\ &\ -0.051
\end{array}
\right),
\end{equation}

\begin{equation} 
t^{d} (0,2a,0) = \left(
\begin{array}{ccccc}
0.005\ &\ 0\ &\ 0\ &\ 0\ &\ 0 \\
0\ &\ 0.022\ &\ 0\ &\ 0\ &\ 0 \\
0\ &\ 0\ &\ 0.001\ &\ 0\ &\ 0 \\
0\ &\ 0\ &\ 0\ &\ -0.036\ &\ 0.023 \\
0\ &\ 0\ &\ 0\ &\ 0.023\ &\ -0.014
\end{array}
\right),
\end{equation}

\begin{equation} 
t^{d} (a,a,0) = \left(
\begin{array}{ccccc}
-0.081\ &\ 0\ &\ 0\ &\ 0.024\ &\ 0.014 \\
0\ &\ 0.006\ &\ 0.000\ &\ 0\ &\ 0 \\
0\ &\ 0.000\ &\ 0.006\ &\ 0\ &\ 0 \\
0.024\ &\ 0\ &\ 0\ &\ 0.004\ &\ -0.025 \\
0.014\ &\ 0\ &\ 0\ &\ -0.025\ &\ 0.033
\end{array}
\right),
\end{equation}

\begin{equation} 
t^{d} (\frac{a}{2},\frac{a}{2},\frac{c}{2}) = \left(
\begin{array}{ccccc}
0.000\ &\ 0.004\ &\ 0.004\ &\ -0.022\ &\ -0.013 \\
0.004\ &\ -0.015\ &\ -0.011\ &\ -0.011\ &\ -0.010 \\
0.004\ &\ -0.011\ &\ -0.015\ &\ -0.014\ &\ -0.004 \\
-0.022\ &\ -0.011\ &\ -0.014\ &\ -0.021\ &\ -0.013 \\
-0.013\ &\ -0.010\ &\ -0.004\ &\ -0.013\ &\ -0.007
\end{array}
\right).
\end{equation}


\subsection{Sr$_2$VO$_3$H}

\begin{equation}
t^{t_{2g}} (0,0,0) = \left(
\begin{array}{ccc}
0\ &\ \ 0\ &\ \ 0 \\
0\ &\ \ 0.393\ &\ \ 0 \\
0\ &\ \ 0\ &\ \ -0.049
\end{array}
\right),
\end{equation}

\begin{equation} 
t^{t_{2g}} (a,0,0) = \left(
\begin{array}{ccc}
0.096\ &\ 0\ &\ 0 \\
0\ &\ -0.056\ &\ 0 \\
0\ &\ 0\ &\ 0.135
\end{array}
\right),
\end{equation}

\begin{equation} 
t^{t_{2g}} (0,b,0) = \left(
\begin{array}{ccc}
-0.442\ &\ 0\ &\ 0 \\
0\ &\ -0.250\ &\ 0 \\
0\ &\ 0\ &\ 0.032
\end{array}
\right),
\end{equation}

\begin{equation} 
t^{t_{2g}} (2a,0,0) = \left(
\begin{array}{ccc}
0.021\ &\ 0\ &\ 0 \\
0\ &\ 0.000\ &\ 0 \\
0\ &\ 0\ &\ 0.003
\end{array}
\right),
\end{equation}

\begin{equation} 
t^{t_{2g}} (0,2b,0) = \left(
\begin{array}{ccc}
-0.005\ &\ 0\ &\ 0 \\
0\ &\ 0.020\ &\ 0 \\
0\ &\ 0\ &\ 0.000
\end{array}
\right),
\end{equation}

\begin{equation} 
t^{t_{2g}} (a,b,0) = \left(
\begin{array}{ccc}
0.025\ &\ 0\ &\ 0 \\
0\ &\ 0.003\ &\ -0.020 \\
0\ &\ -0.020\ &\ -0.029
\end{array}
\right),
\end{equation}

\begin{equation} 
t^{t_{2g}} (\frac{a}{2},\frac{b}{2},\frac{c}{2}) = \left(
\begin{array}{ccc}
0.001\ &\ -0.011\ &\ 0.005 \\
-0.011\ &\ -0.019\ &\ -0.023 \\
0.005\ &\ -0.023\ &\ -0.018
\end{array}
\right).
\end{equation}

For the $d$ model in Sr$_2$VO$_3$H, the orbital index runs as ($d_{xy}$, $d_{yz}$, $d_{xz}$, $d_{y^2-z^2}$, $d_{3x^2-r^2}$),
where hydrogen atoms are aligned along the $x$ direction.
In this model, we also show the imaginary part of the hopping parameters because they become non-negligible amplitude for some matrix elements.

\begin{widetext}
\begin{equation}
t^{d} (0,0,0) = \left(
\begin{array}{ccccc}
0\ &\ 0\ &\ 0\ &\ 0\ &\ 0.001-0.001i \\
0\ &\ 0.378\ &\ 0\ &\ 0\ &\ 0 \\
0\ &\ 0\ &\ -0.069\ &\ 0\ &\ 0.002 \\
0\ &\ 0\ &\ 0\ &\ 2.907\ &\ -0.042-0.042i \\
0.001+0.001i\ &\ 0\ &\ 0.002\ &\ -0.042+0.042i\ &\ 1.901
\end{array}
\right),
\end{equation}

\begin{equation} 
t^{d} (a,0,0) = \left(
\begin{array}{ccccc}
0.091\ &\ 0\ &\ 0\ &\ 0\ &\ 0.001 \\
0\ &\ -0.056\ &\ 0\ &\ 0\ &\ 0 \\
0\ &\ 0\ &\ 0.135\ &\ 0\ &\ 0.001 \\
0\ &\ 0\ &\ 0\ &\ 0\ &\ -0.020-0.039i \\
0.001\ &\ 0\ &\ 0.001\ &\ -0.020+0.039i\ &\ 0.878
\end{array}
\right),
\end{equation}

\begin{equation} 
t^{d} (0,b,0) = \left(
\begin{array}{ccccc}
-0.445\ &\ 0\ &\ 0\ &\ 0\ &\ 0 \\
0\ &\ -0.255\ &\ 0\ &\ 0\ &\ 0 \\
0\ &\ 0\ &\ 0.032\ &\ 0\ &\ 0 \\
0\ &\ 0\ &\ 0\ &\ -0.426\ &\ 0.104+0.133i \\
0\ &\ 0\ &\ 0\ &\ 0.104-0.133i\ &\ -0.095
\end{array}
\right),
\end{equation}

\begin{equation} 
t^{d} (2a,0,0) = \left(
\begin{array}{ccccc}
0.011\ &\ 0\ &\ 0\ &\ 0\ &\ 0 \\
0\ &\ 0\ &\ 0\ &\ 0\ &\ 0 \\
0\ &\ 0\ &\ 0.003\ &\ 0\ &\ 0 \\
0\ &\ 0\ &\ 0\ &\ 0.003\ &\ -0.005-0.006i \\
0\ &\ 0\ &\ 0\ &\ -0.005+0.006i\ &\ -0.055
\end{array}
\right),
\end{equation}

\begin{equation} 
t^{d} (0,2b,0) = \left(
\begin{array}{ccccc}
-0.013\ &\ 0\ &\ 0\ &\ 0\ &\ 0 \\
0\ &\ 0.022\ &\ 0\ &\ 0\ &\ 0 \\
0\ &\ 0\ &\ 0\ &\ 0\ &\ 0 \\
0\ &\ 0\ &\ 0\ &\ -0.040\ &\ 0.012+0.016i \\
0\ &\ 0\ &\ 0\ &\ 0.012-0.016i\ &\ -0.012
\end{array}
\right),
\end{equation}

\begin{equation} 
t^{d} (a,b,0) = \left(
\begin{array}{ccccc}
0.025\ &\ 0\ &\ 0\ &\ -0.005-0.001i\ &\ -0.055-0.072i \\
0\ &\ 0.002\ &\ -0.020\ &\ 0\ &\ 0 \\
0\ &\ -0.020\ &\ -0.029\ &\ 0\ &\ 0 \\
-0.005+0.001i\ &\ 0\ &\ 0\ &\ -0.004\ &\ 0.030+0.042i \\
-0.055+0.072i\ &\ 0\ &\ 0\ &\ 0.030-0.042i\ &\ -0.071
\end{array}
\right),
\end{equation}

\begin{equation} 
t^{d} (\frac{a}{2},\frac{b}{2},\frac{c}{2}) = \left(
\begin{array}{ccccc}
0\ &\ -0.010\ &\ 0.005\ &\ 0.003\ &\ 0.001+0.002i \\
-0.010\ &\ -0.019\ &\ -0.022\ &\ -0.008\ &\ 0.002+0.003i \\
0.005\ &\ -0.022\ &\ -0.018\ &\ -0.005\ &\ -0.005-0.006i \\
0.003\ &\ -0.008\ &\ -0.005\ &\ -0.017\ &\ -0.013-0.017i \\
0.001-0.002i\ &\ 0.002-0.003i\ &\ -0.005+0.006i\ &\ -0.013+0.017i\ &\ -0.021
\end{array}
\right).
\end{equation}
\end{widetext}

\section{Interaction parameters}

The orbital index runs as ($d_{xy}$, $d_{yz}$, $d_{xz}$) for the $t_{2g}$ model,
and ($d_{xy}$, $d_{yz}$, $d_{xz}$, $d_{x^2-y^2}$, $d_{3z^2-r^2}$) for the $d$ model, unless noted.

\subsection{SrVO$_3$}

\begin{equation}
U^{\mathrm{scr}}_{t_{2g}} = \left(
\begin{array}{ccc}
3.42\ \ &\ \ 2.43\ \ &\ \ 2.43 \\
2.43\ \ &\ \ 3.42\ \ &\ \ 2.43 \\
2.43\ \ &\ \ 2.43\ \ &\ \ 3.42
\end{array}
\right),
\end{equation}

\begin{equation}
U^{\mathrm{bare}}_{t_{2g}} = \left(
\begin{array}{ccc}
15.78\ \ &\ \ 14.49\ \ &\ \ 14.49  \\
14.49\ \ &\ \ 15.78\ \ &\ \ 14.49\\
14.49\ \ &\ \ 14.49\ \ &\ \  15.78 
\end{array}
\right),
\end{equation}

\begin{equation}
J^{\mathrm{scr}}_{t_{2g}} = \left(
\begin{array}{ccc}
3.42\ \ &\ \ 0.48\ \ &\ \ 0.48 \\
0.48\ \ &\ \ 3.42\ \ &\ \ 0.48 \\
0.48\ \ &\ \ 0.48\ \ &\ \ 3.42
\end{array}
\right),
\end{equation}

\begin{equation}
J^{\mathrm{bare}}_{t_{2g}} = \left(
\begin{array}{ccc}
15.78\ \ &\ \ 0.61\ \ &\ \ 0.61 \\
0.61\ \ &\ \ 15.78\ \ &\ \ 0.61 \\
0.61\ \ &\ \ 0.61\ \ &\ \ 15.78
\end{array}
\right),
\end{equation}

\begin{equation}
U^{\mathrm{scr}}_{d} = \left(
\begin{array}{ccccc}
3.43 \ \ &\ \  2.44 \ \ &\ \  2.44 \ \ &\ \  2.81 \ \ &\ \  2.37 \\
2.44 \ \ &\ \  3.43 \ \ &\ \  2.44 \ \ &\ \  2.48 \ \ &\ \  2.70 \\
2.44 \ \ &\ \  2.44 \ \ &\ \  3.43 \ \ &\ \  2.48 \ \ &\ \  2.70 \\
2.81 \ \ &\ \  2.48 \ \ &\ \  2.48 \ \ &\ \  3.57 \ \ &\ \  2.43 \\
2.37 \ \ &\ \  2.70 \ \ &\ \  2.70 \ \ &\ \  2.43 \ \ &\ \  3.57
\end{array}
\right),
\end{equation}

\begin{equation}
U^{\mathrm{bare}}_{d} = \left(
\begin{array}{ccccc}
15.85\ \ &\ \  14.55 \ \ &\ \  14.55 \ \ &\ \  15.36 \ \ &\ \  14.52 \\
14.55 \ \ &\ \  15.85 \ \ &\ \  14.55 \ \ &\ \  14.73 \ \ &\ \  15.15 \\
14.55 \ \ &\ \  14.55 \ \ &\ \  15.85 \ \ &\ \  14.73 \ \ &\ \  15.15 \\
15.36 \ \ &\ \  14.73 \ \ &\ \  14.73 \ \ &\ \  16.36 \ \ &\ \  14.73 \\
14.52 \ \ &\ \  15.15 \ \ &\ \  15.15 \ \ &\ \  14.73 \ \ &\ \  16.36
\end{array}
\right),
\end{equation}

\begin{equation}
J^{\mathrm{scr}}_{d} = \left(
\begin{array}{ccccc}
3.43\ \ &\ \  0.48 \ \ &\ \  0.48 \ \ &\ \  0.33 \ \ &\ \  0.53 \\
0.48 \ \ &\ \  3.43 \ \ &\ \  0.48 \ \ &\ \  0.48 \ \ &\ \  0.38 \\
0.48 \ \ &\ \  0.48 \ \ &\ \  3.43 \ \ &\ \  0.48 \ \ &\ \  0.38\\
0.33 \ \ &\ \  0.48 \ \ &\ \  0.48 \ \ &\ \  3.57 \ \ &\ \  0.57\\
0.53 \ \ &\ \  0.38 \ \ &\ \  0.38 \ \ &\ \  0.57 \ \ &\ \  3.57
\end{array}
\right),
\end{equation}

\begin{equation}
J^{\mathrm{bare}}_{d} = \left(
\begin{array}{ccccc}
15.85\ \ &\ \  0.61 \ \ &\ \  0.61 \ \ &\ \  0.35 \ \ &\ \  0.70 \\
0.61 \ \ &\ \  15.85 \ \ &\ \  0.61 \ \ &\ \  0.61 \ \ &\ \  0.44 \\
0.61 \ \ &\ \  0.61 \ \ &\ \  15.85 \ \ &\ \  0.61 \ \ &\ \  0.44 \\
0.35 \ \ &\ \  0.61 \ \ &\ \  0.61 \ \ &\ \  16.36 \ \ &\ \  0.81 \\
0.70 \ \ &\ \  0.44 \ \ &\ \  0.44 \ \ &\ \  0.81 \ \ &\ \  16.36
\end{array}
\right).
\end{equation}

For the $dp$ model, the orbital index runs as
($d_{xy}$, $d_{yz}$, $d_{xz}$, $d_{x^2-y^2}$, $d_{3z^2-r^2}$,
O1-$p_{x,y,z}$, O2-$p_{x,y,z}$, O3-$p_{x,y,z}$),
where O1, O2, and O3 atoms place next to the vanadium atom along the $x$, $y$, and $z$ directions, respectively.

\begin{widetext}
\begin{equation}
U^{\mathrm{scr}}_{dp} = \left(
\begin{array}{ccccc|ccc|ccc|ccc}
11.41 \ \ &\ \  10.01 \ \ &\ \  10.01 \ \ &\ \ 11.02 \ \ &\ \  10.31 \ \ &\ \  3.62 \ \ &\ \  3.08 \ \ &\ \  3.04\ \ &\ \  3.08\ \ &\ \ 3.62 \ \ &\ \  3.04\ \ &\ \  2.87 \ \ &\ \  2.87\ \ &\ \  3.32 \\
10.01 \ \ &\ \  11.41 \ \ &\ \  10.01 \ \ &\ \ 10.49 \ \ &\ \  10.84 \ \ &\ \  3.32 \ \ &\ \  2.87 \ \ &\ \  2.87\ \ &\ \  3.04\ \ &\ \ 3.62 \ \ &\ \  3.08\ \ &\ \  3.04 \ \ &\ \  3.08\ \ &\ \  3.62 \\
10.01 \ \ &\ \  10.01 \ \ &\ \  11.41 \ \ &\ \ 10.49 \ \ &\ \  10.84 \ \ &\ \  3.62 \ \ &\ \  3.04 \ \ &\ \  3.08\ \ &\ \  2.87\ \ &\ \ 3.32 \ \ &\ \  2.87\ \ &\ \  3.08 \ \ &\ \  3.04\ \ &\ \  3.62 \\
11.02 \ \ &\ \  10.49 \ \ &\ \  10.49 \ \ &\ \ 12.64 \ \ &\ \  10.83 \ \ &\ \  3.68 \ \ &\ \  3.10 \ \ &\ \  3.08\ \ &\ \  3.10\ \ &\ \ 3.68 \ \ &\ \  3.08\ \ &\ \  2.89 \ \ &\ \  2.89\ \ &\ \  3.37 \\
10.31 \ \ &\ \  10.84 \ \ &\ \  10.84 \ \ &\ \ 10.83 \ \ &\ \  12.64 \ \ &\ \  3.47 \ \ &\ \  2.95 \ \ &\ \  2.96\ \ &\ \  2.95\ \ &\ \ 3.47 \ \ &\ \  2.96\ \ &\ \  3.16 \ \ &\ \  3.16\ \ &\ \  3.79 \\
\hline
3.62 \ \ &\ \  3.32 \ \ &\ \  3.62 \ \ &\ \ 3.68 \ \ &\ \  3.47 \ \ &\ \  8.73 \ \ &\ \  6.81 \ \ &\ \  6.81 \ \ &\ \  2.10 \ \ &\ \ 2.14 \ \ &\ \  2.01 \ \ &\ \  2.10 \ \ &\ \  2.01 \ \ &\ \  2.14 \\
3.08 \ \ &\ \  2.87 \ \ &\ \  3.04 \ \ &\ \ 3.10 \ \ &\ \  2.95 \ \ &\ \  6.81 \ \ &\ \  8.07 \ \ &\ \  6.55 \ \ &\ \  2.12 \ \ &\ \ 2.10 \ \ &\ \  2.00 \ \ &\ \  2.00 \ \ &\ \  1.93 \ \ &\ \  2.01 \\
3.04 \ \ &\ \  2.87 \ \ &\ \  3.08 \ \ &\ \ 3.08 \ \ &\ \  2.96 \ \ &\ \  6.81 \ \ &\ \  6.55 \ \ &\ \  8.07 \ \ &\ \  2.00 \ \ &\ \ 2.01 \ \ &\ \  1.93 \ \ &\ \  2.12 \ \ &\ \  2.00 \ \ &\ \  2.10 \\
\hline
3.08 \ \ &\ \  3.04 \ \ &\ \  2.87 \ \ &\ \ 3.10 \ \ &\ \  2.95 \ \ &\ \  2.10 \ \ &\ \  2.12 \ \ &\ \  2.00 \ \ &\ \  8.07 \ \ &\ \ 6.81 \ \ &\ \  6.55 \ \ &\ \  1.93 \ \ &\ \  2.00 \ \ &\ \  2.01 \\
3.62 \ \ &\ \  3.62 \ \ &\ \  3.32 \ \ &\ \ 3.68 \ \ &\ \  3.47 \ \ &\ \  2.14 \ \ &\ \  2.10 \ \ &\ \  2.01 \ \ &\ \  6.81 \ \ &\ \ 8.73 \ \ &\ \  6.81 \ \ &\ \  2.01 \ \ &\ \  2.10 \ \ &\ \  2.14 \\
3.04 \ \ &\ \  3.08 \ \ &\ \  2.87 \ \ &\ \ 3.08 \ \ &\ \  2.96 \ \ &\ \  2.01 \ \ &\ \  2.00 \ \ &\ \  1.93 \ \ &\ \  6.55 \ \ &\ \ 6.81 \ \ &\ \  8.07 \ \ &\ \  2.00 \ \ &\ \  2.12 \ \ &\ \  2.10 \\
\hline
2.87 \ \ &\ \  3.04 \ \ &\ \  3.08 \ \ &\ \ 2.89 \ \ &\ \  3.16 \ \ &\ \  2.10 \ \ &\ \  2.00 \ \ &\ \  2.12 \ \ &\ \  1.93 \ \ &\ \ 2.01 \ \ &\ \  2.00 \ \ &\ \  8.07 \ \ &\ \  6.55 \ \ &\ \  6.81 \\
2.87 \ \ &\ \  3.08 \ \ &\ \  3.04 \ \ &\ \ 2.89 \ \ &\ \  3.16 \ \ &\ \  2.01 \ \ &\ \  1.93 \ \ &\ \  2.00 \ \ &\ \  2.00 \ \ &\ \ 2.10 \ \ &\ \  2.12 \ \ &\ \  6.55 \ \ &\ \  8.07 \ \ &\ \  6.81 \\
3.32 \ \ &\ \  3.62 \ \ &\ \  3.62 \ \ &\ \ 3.37 \ \ &\ \  3.79 \ \ &\ \  2.14 \ \ &\ \  2.01 \ \ &\ \  2.10 \ \ &\ \  2.01 \ \ &\ \ 2.14 \ \ &\ \  2.10 \ \ &\ \  6.81 \ \ &\ \  6.81 \ \ &\ \  8.73
\end{array}
\right),
\end{equation}
\end{widetext}

\begin{widetext}
\begin{equation}
U^{\mathrm{bare}}_{dp} = \left(
\begin{array}{ccccc|ccc|ccc|ccc}
19.36 \ \ &\ \  17.72 \ \ &\ \  17.72 \ \ &\ \ 19.28 \ \ &\ \  18.28\ \ &\ \  7.88\ \ &\ \  7.10 \ \ &\ \  7.04\ \ &\ \  7.10\ \ &\ \ 7.88 \ \ &\ \  7.04\ \ &\ \  6.72 \ \ &\ \  6.72 \ \ &\ \  7.39 \\
17.72 \ \ &\ \  19.36 \ \ &\ \  17.72 \ \ &\ \ 18.53 \ \ &\ \  19.03\ \ &\ \  7.39\ \ &\ \  6.72 \ \ &\ \  6.72\ \ &\ \  7.04\ \ &\ \ 7.88 \ \ &\ \  7.10\ \ &\ \  7.04 \ \ &\ \  7.10 \ \ &\ \  7.88 \\
17.72 \ \ &\ \  17.72 \ \ &\ \  19.36 \ \ &\ \ 18.53 \ \ &\ \  19.03\ \ &\ \  7.88\ \ &\ \  7.04 \ \ &\ \  7.10\ \ &\ \  6.72\ \ &\ \ 7.39 \ \ &\ \  6.72\ \ &\ \  7.10 \ \ &\ \  7.04 \ \ &\ \  7.88 \\
19.28 \ \ &\ \  18.53 \ \ &\ \  18.53 \ \ &\ \ 21.31 \ \ &\ \  19.16\ \ &\ \  8.05\ \ &\ \  7.17 \ \ &\ \  7.15\ \ &\ \  7.17\ \ &\ \ 8.05 \ \ &\ \  7.15\ \ &\ \  6.75 \ \ &\ \  6.75 \ \ &\ \  7.46 \\
18.28 \ \ &\ \  19.03 \ \ &\ \  19.03 \ \ &\ \ 19.16 \ \ &\ \  21.31\ \ &\ \  7.66\ \ &\ \  6.88 \ \ &\ \  6.90\ \ &\ \  6.88\ \ &\ \ 7.66 \ \ &\ \  6.90\ \ &\ \  7.30 \ \ &\ \  7.30 \ \ &\ \  8.24 \\
\hline
7.88 \ \ &\ \  7.39 \ \ &\ \  7.88 \ \ &\ \ 8.05 \ \ &\ \  7.66\ \ &\ \  19.82\ \ &\ \  17.24 \ \ &\ \  17.24\ \ &\ \  5.23\ \ &\ \ 5.30 \ \ &\ \  5.08\ \ &\ \  5.23 \ \ &\ \  5.08 \ \ &\ \  5.30 \\
7.10 \ \ &\ \  6.72 \ \ &\ \  7.04 \ \ &\ \ 7.17 \ \ &\ \  6.88\ \ &\ \  17.24\ \ &\ \  18.55 \ \ &\ \  16.71\ \ &\ \  5.24\ \ &\ \ 5.23 \ \ &\ \  5.06\ \ &\ \  5.06 \ \ &\ \  4.93 \ \ &\ \  5.08 \\
7.04 \ \ &\ \  6.72 \ \ &\ \  7.10 \ \ &\ \ 7.15 \ \ &\ \  6.90\ \ &\ \  17.24\ \ &\ \  16.71 \ \ &\ \  18.55\ \ &\ \  5.06\ \ &\ \ 5.08 \ \ &\ \  4.93\ \ &\ \  5.24 \ \ &\ \  5.06 \ \ &\ \  5.23 \\
\hline
7.10 \ \ &\ \  7.04 \ \ &\ \  6.72 \ \ &\ \ 7.17 \ \ &\ \  6.88\ \ &\ \  5.23\ \ &\ \  5.24 \ \ &\ \  5.06\ \ &\ \  18.55\ \ &\ \ 17.24 \ \ &\ \  16.71\ \ &\ \  4.93 \ \ &\ \  5.06 \ \ &\ \  5.08 \\
7.88 \ \ &\ \  7.88 \ \ &\ \  7.39 \ \ &\ \ 8.05 \ \ &\ \  7.66\ \ &\ \  5.30\ \ &\ \  5.23 \ \ &\ \  5.08\ \ &\ \  17.24\ \ &\ \ 19.82 \ \ &\ \  17.24\ \ &\ \  5.08 \ \ &\ \  5.23 \ \ &\ \  5.30 \\
7.04 \ \ &\ \  7.10 \ \ &\ \  6.72 \ \ &\ \ 7.15 \ \ &\ \  6.90\ \ &\ \  5.08\ \ &\ \  5.06 \ \ &\ \  4.93\ \ &\ \  16.71\ \ &\ \ 17.24 \ \ &\ \  18.55\ \ &\ \  5.06 \ \ &\ \  5.24 \ \ &\ \  5.23 \\
\hline
6.72 \ \ &\ \  7.04 \ \ &\ \  7.10 \ \ &\ \ 6.75 \ \ &\ \  7.30\ \ &\ \  5.23\ \ &\ \  5.06 \ \ &\ \  5.24\ \ &\ \  4.93\ \ &\ \ 5.08 \ \ &\ \  5.06\ \ &\ \  18.55 \ \ &\ \  16.71\ \ &\ \  17.24 \\
6.72 \ \ &\ \  7.10 \ \ &\ \  7.04 \ \ &\ \ 6.75 \ \ &\ \  7.30\ \ &\ \  5.08\ \ &\ \  4.93 \ \ &\ \  5.06\ \ &\ \  5.06\ \ &\ \ 5.23 \ \ &\ \  5.24\ \ &\ \  16.71 \ \ &\ \  18.55 \ \ &\ \  17.24 \\
7.39 \ \ &\ \  7.88 \ \ &\ \  7.88 \ \ &\ \ 7.46 \ \ &\ \  8.24\ \ &\ \  5.30\ \ &\ \  5.08 \ \ &\ \  5.23\ \ &\ \  5.08\ \ &\ \ 5.30 \ \ &\ \  5.23\ \ &\ \  17.24 \ \ &\ \  17.24 \ \ &\ \  19.82
\end{array}
\right).
\end{equation}
\end{widetext}

Because the off-site exchange interaction is less than 0.1 eV, we only show the on-site exchange terms here.
For the V-$d$ orbitals,

\begin{equation}
J^{\mathrm{scr},d}_{dp} = \left(
\begin{array}{ccccc}
11.41 \ \ &\ \  0.71 \ \ &\ \  0.71 \ \ &\ \ 0.48 \ \ &\ \  0.85 \\
0.71 \ \ &\ \  11.41 \ \ &\ \  0.71 \ \ &\ \ 0.76 \ \ &\ \  0.57 \\
0.71 \ \ &\ \  0.71 \ \ &\ \  11.41 \ \ &\ \ 0.76 \ \ &\ \  0.57 \\
0.48 \ \ &\ \  0.76 \ \ &\ \  0.76 \ \ &\ \ 12.64 \ \ &\ \  0.91 \\
0.85 \ \ &\ \  0.57 \ \ &\ \  0.57 \ \ &\ \ 0.91 \ \ &\ \  12.64 \\
\end{array}
\right),
\end{equation}

\begin{equation}
J^{\mathrm{bare},d}_{dp} = \left(
\begin{array}{ccccc}
19.36 \ \ &\ \  0.82 \ \ &\ \  0.82 \ \ &\ \ 0.49 \ \ &\ \  1.00 \\
0.82 \ \ &\ \  19.36 \ \ &\ \  0.82 \ \ &\ \ 0.87 \ \ &\ \  0.62 \\
0.82 \ \ &\ \  0.82 \ \ &\ \  19.36 \ \ &\ \ 0.87 \ \ &\ \  0.62 \\
0.49 \ \ &\ \  0.87 \ \ &\ \  0.87 \ \ &\ \ 21.31 \ \ &\ \  1.08 \\
1.00 \ \ &\ \  0.62 \ \ &\ \  0.62 \ \ &\ \ 1.08 \ \ &\ \  21.31 \\
\end{array}
\right).
\end{equation}

For the O1-$p$ orbitals,

\begin{equation}
J^{\mathrm{scr},p}_{dp} = \left(
\begin{array}{ccc}
8.73\ \ &\ \  0.80 \ \ &\ \  0.80 \\
0.80\ \ &\ \  8.07 \ \ &\ \  0.76 \\
0.80\ \ &\ \  0.76 \ \ &\ \  8.07 \\
\end{array}
\right),
\end{equation}

\begin{equation}
J^{\mathrm{bare},p}_{dp} = \left(
\begin{array}{ccc}
19.82\ \ &\ \ 0.97 \ \ &\ \  0.97 \\
0.97\ \ &\ \  18.55 \ \ &\ \  0.93 \\
0.97\ \ &\ \  0.93 \ \ &\ \  18.55\ \\
\end{array}
\right).
\end{equation}

The on-site exchange terms for the O2 and O3 atoms are equivalent to the above ones by appropriately exchanging the orbital indices.

\subsection{SrVO$_2$H}

\begin{equation}
U^{\mathrm{scr}}_{t_{2g}} = \left(
\begin{array}{ccccc}
3.00 \ \ &\ \  1.85 \ \ &\ \  1.85 \\
1.85 \ \ &\ \  2.60 \ \ &\ \  1.69 \\
1.85 \ \ &\ \  1.69 \ \ &\ \  2.60
\end{array}
\right),
\end{equation}

\begin{equation}
U^{\mathrm{bare}}_{t_{2g}} = \left(
\begin{array}{ccccc}
16.04 \ \ &\ \  14.36 \ \ &\ \  14.36 \\
14.36 \ \ &\ \  15.18 \ \ &\ \  13.91 \\
14.36 \ \ &\ \  13.91 \ \ &\ \  15.18
\end{array}
\right),
\end{equation}

\begin{equation}
J^{\mathrm{scr}}_{t_{2g}} = \left(
\begin{array}{ccccc}
3.00 \ \ &\ \  0.46 \ \ &\ \  0.46 \\
0.46 \ \ &\ \  2.60 \ \ &\ \  0.42\\
0.46 \ \ &\ \  0.42 \ \ &\ \  2.60
\end{array}
\right),
\end{equation}

\begin{equation}
J^{\mathrm{bare}}_{t_{2g}} = \left(
\begin{array}{ccccc}
16.04 \ \ &\ \  0.60 \ \ &\ \  0.60 \\
0.60 \ \ &\ \  15.18 \ \ &\ \  0.57 \\
0.60 \ \ &\ \  0.57 \ \ &\ \  15.18
\end{array}
\right),
\end{equation}

\begin{equation}
U^{\mathrm{scr}}_{d} = \left(
\begin{array}{ccccc}
3.97 \ \ &\ \  2.88 \ \ &\ \  2.88 \ \ &\ \ 3.30 \ \ &\ \  2.55\\
2.88 \ \ &\ \  3.75 \ \ &\ \  2.79 \ \ &\ \ 2.88 \ \ &\ \  2.78\\
2.88 \ \ &\ \  2.79 \ \ &\ \  3.75 \ \ &\ \ 2.88 \ \ &\ \  2.78\\
3.30 \ \ &\ \  2.88 \ \ &\ \  2.88 \ \ &\ \ 4.04 \ \ &\ \  2.57\\
2.55 \ \ &\ \  2.78 \ \ &\ \  2.78 \ \ &\ \ 2.57 \ \ &\ \  3.26
\end{array}
\right),
\end{equation}

\begin{equation}
U^{\mathrm{bare}}_{d} = \left(
\begin{array}{ccccc}
16.06 \ \ &\ \  14.41 \ \ &\ \  14.41 \ \ &\ \ 15.46 \ \ &\ \  13.35\\
14.41 \ \ &\ \  15.28 \ \ &\ \  13.99 \ \ &\ \ 14.50 \ \ &\ \  13.52\\
14.41 \ \ &\ \  13.99 \ \ &\ \  15.28 \ \ &\ \ 14.50 \ \ &\ \  13.52\\
15.46 \ \ &\ \  14.50 \ \ &\ \  14.50 \ \ &\ \ 16.37 \ \ &\ \  13.44\\
13.35 \ \ &\ \  13.52 \ \ &\ \  13.52 \ \ &\ \ 13.44 \ \ &\ \  13.58
\end{array}
\right),
\end{equation}

\begin{equation}
J^{\mathrm{scr}}_{d} = \left(
\begin{array}{ccccc}
3.97 \ \ &\ \  0.49 \ \ &\ \  0.49 \ \ &\ \ 0.34 \ \ &\ \  0.47\\
0.49 \ \ &\ \  3.75 \ \ &\ \  0.45 \ \ &\ \ 0.47 \ \ &\ \  0.32\\
0.49 \ \ &\ \  0.45 \ \ &\ \  3.75 \ \ &\ \ 0.47 \ \ &\ \  0.32\\
0.34 \ \ &\ \  0.47 \ \ &\ \  0.47 \ \ &\ \ 4.04 \ \ &\ \  0.51\\
0.47 \ \ &\ \  0.32 \ \ &\ \  0.32 \ \ &\ \ 0.51 \ \ &\ \  3.26
\end{array}
\right),
\end{equation}

\begin{equation}
J^{\mathrm{bare}}_{d} = \left(
\begin{array}{ccccc}
16.06 \ \ &\ \  0.60 \ \ &\ \  0.60 \ \ &\ \ 0.36 \ \ &\ \  0.62\\
0.60 \ \ &\ \  15.28 \ \ &\ \  0.57 \ \ &\ \ 0.59 \ \ &\ \  0.38\\
0.60 \ \ &\ \  0.57 \ \ &\ \  15.28 \ \ &\ \ 0.59 \ \ &\ \  0.38\\
0.36 \ \ &\ \  0.59 \ \ &\ \  0.59 \ \ &\ \ 16.37 \ \ &\ \  0.69\\
0.62 \ \ &\ \  0.38 \ \ &\ \  0.38 \ \ &\ \ 0.69 \ \ &\ \  13.58
\end{array}
\right).
\end{equation}

For the $dps$ model, the orbital index runs as
($d_{xy}$, $d_{yz}$, $d_{xz}$, $d_{x^2-y^2}$, $d_{3z^2-r^2}$,
O1-$p_{x,y,z}$, O2-$p_{x,y,z}$, H-$s$),
where O1, O2, and H atoms place next to the vanadium atom along the $x$, $y$, and $z$ directions, respectively.

\begin{widetext}
\begin{equation}
U^{\mathrm{scr}}_{dps} = \left(
\begin{array}{ccccc|ccc|ccc|c}
8.16 \ \ &\ \  6.48 \ \ &\ \ 6.48 \ \ &\ \ 7.61 \ \ &\ \  6.53\ \ &\ \  2.99\ \ &\ \  2.63\ \ &\ \  2.61\ \ &\ \  2.63\ \ &\ \ 2.99\ \ &\ \  2.61\ \ &\ \  2.83 \\
6.48 \ \ &\ \  7.28 \ \ &\ \ 6.17 \ \ &\ \ 6.72 \ \ &\ \  6.65\ \ &\ \  2.66\ \ &\ \  2.38\ \ &\ \  2.40\ \ &\ \  2.56\ \ &\ \ 2.94\ \ &\ \  2.63\ \ &\ \  3.08 \\
6.48 \ \ &\ \  6.17 \ \ &\ \ 7.28 \ \ &\ \ 6.72 \ \ &\ \  6.65\ \ &\ \  2.94\ \ &\ \  2.56\ \ &\ \  2.63\ \ &\ \  2.38\ \ &\ \ 2.66\ \ &\ \  2.40\ \ &\ \  3.08 \\
7.61 \ \ &\ \  6.72 \ \ &\ \ 6.72 \ \ &\ \ 8.95 \ \ &\ \  6.79\ \ &\ \  3.03\ \ &\ \  2.62\ \ &\ \  2.63\ \ &\ \  2.62\ \ &\ \ 3.03\ \ &\ \  2.63\ \ &\ \  2.86 \\
6.53 \ \ &\ \  6.65 \ \ &\ \ 6.65 \ \ &\ \ 6.79 \ \ &\ \  7.90\ \ &\ \  2.74\ \ &\ \  2.41\ \ &\ \  2.45\ \ &\ \  2.41\ \ &\ \ 2.74\ \ &\ \  2.45\ \ &\ \  3.28 \\
\hline
2.99 \ \ &\ \  2.66 \ \ &\ \ 2.94 \ \ &\ \ 3.03 \ \ &\ \  2.74\ \ &\ \  7.60\ \ &\ \  6.02\ \ &\ \  6.07\ \ &\ \  1.85\ \ &\ \ 1.87\ \ &\ \  1.78\ \ &\ \  1.91 \\
2.63 \ \ &\ \  2.38 \ \ &\ \ 2.56 \ \ &\ \ 2.62 \ \ &\ \  2.41\ \ &\ \  6.02\ \ &\ \  7.44\ \ &\ \  6.01\ \ &\ \  1.87\ \ &\ \ 1.85\ \ &\ \  1.77\ \ &\ \  1.81 \\
2.61 \ \ &\ \  2.40 \ \ &\ \ 2.63 \ \ &\ \ 2.63 \ \ &\ \  2.45\ \ &\ \  6.07\ \ &\ \  6.01\ \ &\ \  7.58\ \ &\ \  1.77\ \ &\ \ 1.78 \ \ &\ \  1.71\ \ &\ \  1.89 \\
\hline
2.63 \ \ &\ \  2.56 \ \ &\ \ 2.38 \ \ &\ \ 2.62 \ \ &\ \  2.41\ \ &\ \  1.85\ \ &\ \  1.87 \ \ &\ \  1.77\ \ &\ \  7.44\ \ &\ \ 6.02 \ \ &\ \  6.01\ \ &\ \  1.81 \\
2.99 \ \ &\ \  2.94 \ \ &\ \ 2.66 \ \ &\ \ 3.03 \ \ &\ \  2.74\ \ &\ \  1.87\ \ &\ \  1.85 \ \ &\ \  1.78\ \ &\ \  6.02\ \ &\ \ 7.60 \ \ &\ \  6.07\ \ &\ \  1.91 \\
2.61 \ \ &\ \  2.63 \ \ &\ \ 2.40 \ \ &\ \ 2.63 \ \ &\ \  2.45\ \ &\ \  1.78\ \ &\ \  1.77 \ \ &\ \  1.71\ \ &\ \  6.01\ \ &\ \ 6.07 \ \ &\ \  7.58\ \ &\ \  1.89 \\
\hline
2.83 \ \ &\ \  3.08 \ \ &\ \ 3.08 \ \ &\ \ 2.86 \ \ &\ \  3.28\ \ &\ \  1.91\ \ &\ \  1.81 \ \ &\ \  1.89\ \ &\ \  1.81\ \ &\ \ 1.91 \ \ &\ \  1.89\ \ &\ \  6.44
\end{array}
\right),
\end{equation}
\end{widetext}

\begin{widetext}
\begin{equation}
U^{\mathrm{bare}}_{dps} = \left(
\begin{array}{ccccc|ccc|ccc|c}
18.59 \ \ &\ \  16.32 \ \ &\ \  16.32 \ \ &\ \ 18.57 \ \ &\ \  16.59\ \ &\ \  7.64\ \ &\ \  6.92 \ \ &\ \  6.88\ \ &\ \  6.92\ \ &\ \ 7.64 \ \ &\ \  6.88\ \ &\ \  7.06 \\
16.32 \ \ &\ \  17.02 \ \ &\ \  15.67 \ \ &\ \ 17.06 \ \ &\ \  16.58\ \ &\ \  7.06\ \ &\ \  6.46 \ \ &\ \  6.48\ \ &\ \  6.80\ \ &\ \ 7.55 \ \ &\ \  6.89\ \ &\ \  7.44 \\
16.32 \ \ &\ \  15.67 \ \ &\ \  17.02 \ \ &\ \ 17.06 \ \ &\ \  16.58\ \ &\ \  7.55\ \ &\ \  6.80 \ \ &\ \  6.89\ \ &\ \  6.46\ \ &\ \ 7.06 \ \ &\ \  6.48\ \ &\ \  7.44 \\
18.57 \ \ &\ \  17.06 \ \ &\ \  17.06 \ \ &\ \ 20.55 \ \ &\ \  17.39\ \ &\ \  7.79\ \ &\ \  6.97 \ \ &\ \  6.97\ \ &\ \  6.97\ \ &\ \ 7.79 \ \ &\ \  6.97\ \ &\ \  7.11 \\
16.59 \ \ &\ \  16.58 \ \ &\ \  16.58 \ \ &\ \ 17.39 \ \ &\ \  18.26\ \ &\ \  7.20\ \ &\ \  6.51 \ \ &\ \  6.57\ \ &\ \  6.51\ \ &\ \ 7.20 \ \ &\ \  6.57\ \ &\ \  7.77 \\
\hline
7.64 \ \ &\ \  7.06 \ \ &\ \  7.55 \ \ &\ \ 7.79 \ \ &\ \  7.20\ \ &\ \  19.03\ \ &\ \  16.82 \ \ &\ \  16.91\ \ &\ \  5.10\ \ &\ \ 5.16 \ \ &\ \  4.97\ \ &\ \  5.22 \\
6.92 \ \ &\ \  6.46 \ \ &\ \  6.80 \ \ &\ \ 6.97 \ \ &\ \  6.51\ \ &\ \  16.82\ \ &\ \  18.36 \ \ &\ \  16.62\ \ &\ \  5.12\ \ &\ \ 5.10 \ \ &\ \  4.95\ \ &\ \  5.04 \\
6.88 \ \ &\ \  6.48 \ \ &\ \  6.89 \ \ &\ \ 6.97 \ \ &\ \  6.57\ \ &\ \  16.91\ \ &\ \  16.62 \ \ &\ \  18.54\ \ &\ \  4.95\ \ &\ \ 4.97 \ \ &\ \  4.83\ \ &\ \  5.18 \\
\hline
6.92 \ \ &\ \  6.80 \ \ &\ \  6.46 \ \ &\ \ 6.97 \ \ &\ \  6.51\ \ &\ \  5.10\ \ &\ \  5.12 \ \ &\ \  4.95\ \ &\ \  18.36\ \ &\ \ 16.82 \ \ &\ \  16.62\ \ &\ \  5.04 \\
7.64 \ \ &\ \  7.55 \ \ &\ \  7.06 \ \ &\ \ 7.79 \ \ &\ \  7.20\ \ &\ \  5.16\ \ &\ \  5.10 \ \ &\ \  4.97\ \ &\ \  16.82\ \ &\ \ 19.03 \ \ &\ \  16.91\ \ &\ \  5.22 \\
6.88 \ \ &\ \  6.89 \ \ &\ \  6.48 \ \ &\ \ 6.97 \ \ &\ \  6.57\ \ &\ \  4.97\ \ &\ \  4.95 \ \ &\ \  4.83\ \ &\ \  16.62\ \ &\ \ 16.91 \ \ &\ \  18.54\ \ &\ \  5.18 \\
\hline
7.06 \ \ &\ \  7.44 \ \ &\ \  7.44 \ \ &\ \ 7.11 \ \ &\ \  7.77\ \ &\ \  5.22\ \ &\ \  5.04 \ \ &\ \  5.18\ \ &\ \  5.04\ \ &\ \ 5.22 \ \ &\ \  5.18\ \ &\ \  14.14
\end{array}
\right).
\end{equation}
\end{widetext}

Because the off-site exchange interaction is less than 0.1 eV except that between the $d_{3z^2-r^2}$ and $s$ orbitals, 
we only show the $d$-$d$, $d$-$s$, and $p$-$p$ exchange terms here.
For the V-$d$ $+$ H-$s$ orbitals,

\begin{equation}
J^{\mathrm{scr},ds}_{dps} = \left(
\begin{array}{ccccc|c}
8.16 \ \ &\ \  0.62 \ \ &\ \  0.62 \ \ &\ \ 0.45 \ \ &\ \ 0.69 \ \ &\ \  0.01\\
0.62 \ \ &\ \  7.28 \ \ &\ \  0.57 \ \ &\ \ 0.64 \ \ &\ \ 0.47 \ \ &\ \  0.09\\
0.62 \ \ &\ \  0.57 \ \ &\ \  7.28 \ \ &\ \ 0.64 \ \ &\ \ 0.47 \ \ &\ \  0.09\\
0.45 \ \ &\ \  0.64 \ \ &\ \  0.64 \ \ &\ \ 8.95 \ \ &\ \ 0.76 \ \ &\ \  0.01\\
0.69 \ \ &\ \  0.47 \ \ &\ \  0.47 \ \ &\ \ 0.76 \ \ &\ \ 7.90 \ \ &\ \  0.16\\
\hline
0.01 \ \ &\ \  0.09 \ \ &\ \  0.09 \ \ &\ \ 0.01 \ \ &\ \ 0.16 \ \ &\ \  6.44
\end{array}
\right),
\end{equation}

\begin{equation}
J^{\mathrm{bare},ds}_{dps} = \left(
\begin{array}{ccccc|c}
18.59 \ \ &\  0.73 \ \ &\  0.73 \ \ &\ 0.47 \ \ &\  0.84\ \ &\ 0.02\\
0.73 \ \ &\  17.02 \ \ &\  0.69 \ \ &\ 0.77 \ \ &\  0.53\ \ &\ 0.15\\
0.73 \ \ &\  0.69 \ \ &\  17.02 \ \ &\ 0.77 \ \ &\  0.53\ \ &\ 0.15\\
0.47 \ \ &\  0.77 \ \ &\  0.77 \ \ &\ 20.55 \ \ &\  0.92\ \ &\ 0.01\\
0.84 \ \ &\  0.53 \ \ &\  0.53 \ \ &\ 0.92 \ \ &\  18.26\ \ &\ 0.25\\
\hline
0.02 \ \ &\ 0.15 \ \ &\ 0.15 \ \ &\ 0.01 \ \ &\  0.25\ \ &\ 14.14
\end{array}
\right).
\end{equation}

For the O1-$p$ orbitals,

\begin{equation}
J^{\mathrm{scr},p}_{dps} = \left(
\begin{array}{ccc}
7.60 \ \ &\ \ 0.76 \ \ &\ \  0.76\\
0.76 \ \ &\ \ 7.44 \ \ &\ \  0.75\\
0.76 \ \ &\ \ 0.75 \ \ &\ \  7.58
\end{array}
\right),
\end{equation}

\begin{equation}
J^{\mathrm{bare},p}_{dps} = \left(
\begin{array}{ccc}
19.03 \ \ &\ \ 0.94 \ \ &\ \  0.94\\
0.94 \ \ &\ \ 18.36 \ \ &\ \  0.92\\
0.94 \ \ &\ \ 0.92 \ \ &\ \  18.54
\end{array}
\right).
\end{equation}

\subsection{SrCrO$_3$}

\begin{equation}
U^{\mathrm{scr}}_{t_{2g}} = \left(
\begin{array}{ccc}
2.97\ \ &\ \ 2.00\ \ &\ \ 2.00 \\
2.00\ \ &\ \ 2.97\ \ &\ \ 2.00 \\
2.00\ \ &\ \ 2.00\ \ &\ \ 2.97 
\end{array}
\right),
\end{equation}

\begin{equation}
U^{\mathrm{bare}}_{t_{2g}} = \left(
\begin{array}{ccc}
16.18\ \ &\ \ 14.89\ \ &\ \ 14.89 \\
14.89\ \ &\ \ 16.18\ \ &\ \ 14.89 \\
14.89\ \ &\ \ 14.89\ \ &\ \  16.18 
\end{array}
\right),
\end{equation}

\begin{equation}
J^{\mathrm{scr}}_{t_{2g}} = \left(
\begin{array}{ccc}
2.97\ \ &\ \ 0.45\ \ &\ \ 0.45 \\
0.45\ \ &\ \ 2.97\ \ &\ \ 0.45 \\
0.45\ \ &\ \ 0.45\ \ &\ \ 2.97
\end{array}
\right),
\end{equation}

\begin{equation}
J^{\mathrm{bare}}_{t_{2g}} = \left(
\begin{array}{ccc}
16.18\ \ &\ \ 0.59\ \ &\ \ 0.59 \\
0.59\ \ &\ \ 16.18\ \ &\ \ 0.59 \\
0.59\ \ &\ \ 0.59\ \ &\ \ 16.18
\end{array}
\right),
\end{equation}

\begin{equation}
U^{\mathrm{scr}}_{d} = \left(
\begin{array}{ccccc}
3.04 \ \ &\ \  2.07 \ \ &\ \  2.07 \ \ &\ \  2.43 \ \ &\ \  1.99 \\
2.07 \ \ &\ \  3.04 \ \ &\ \  2.07 \ \ &\ \  2.10 \ \ &\ \  2.32 \\
2.07 \ \ &\ \  2.07 \ \ &\ \  3.04 \ \ &\ \  2.10 \ \ &\ \  2.32 \\
2.43 \ \ &\ \  2.10 \ \ &\ \  2.10 \ \ &\ \  3.18 \ \ &\ \  2.05 \\
1.99 \ \ &\ \  2.32 \ \ &\ \  2.32 \ \ &\ \  2.05 \ \ &\ \  3.18
\end{array}
\right),
\end{equation}

\begin{equation}
U^{\mathrm{bare}}_{d} = \left(
\begin{array}{ccccc}
16.20 \ \ &\ \  14.91 \ \ &\ \  14.91 \ \ &\ \  15.78 \ \ &\ \  14.93 \\
14.91 \ \ &\ \  16.20 \ \ &\ \  14.91 \ \ &\ \  15.14 \ \ &\ \  15.57 \\
14.91 \ \ &\ \  14.91 \ \ &\ \  16.20 \ \ &\ \  15.14 \ \ &\ \  15.57 \\
15.78 \ \ &\ \  15.14 \ \ &\ \  15.14 \ \ &\ \  16.82 \ \ &\ \  15.19 \\
14.93 \ \ &\ \  15.57 \ \ &\ \  15.57 \ \ &\ \  15.19 \ \ &\ \  16.82
\end{array}
\right),
\end{equation}

\begin{equation}
J^{\mathrm{scr}}_{d} = \left(
\begin{array}{ccccc}
3.04 \ \ &\ \  0.47 \ \ &\ \  0.47 \ \ &\ \  0.33 \ \ &\ \  0.52 \\
0.47 \ \ &\ \  3.04 \ \ &\ \  0.47 \ \ &\ \  0.47 \ \ &\ \  0.38 \\
0.47 \ \ &\ \  0.47 \ \ &\ \  3.04 \ \ &\ \  0.47 \ \ &\ \  0.38 \\
0.33 \ \ &\ \  0.47 \ \ &\ \  0.47 \ \ &\ \  3.18 \ \ &\ \  0.56 \\
0.52 \ \ &\ \  0.38 \ \ &\ \  0.38 \ \ &\ \  0.56 \ \ &\ \  3.18
\end{array}
\right),
\end{equation}

\begin{equation}
J^{\mathrm{bare}}_{d} = \left(
\begin{array}{ccccc}
16.20 \ \ &\ \  0.60 \ \ &\ \  0.60 \ \ &\ \  0.35 \ \ &\ \  0.68 \\
0.60 \ \ &\ \  16.20 \ \ &\ \  0.60 \ \ &\ \  0.60 \ \ &\ \  0.43 \\
0.60 \ \ &\ \  0.60 \ \ &\ \  16.20 \ \ &\ \  0.60 \ \ &\ \  0.43 \\
0.35 \ \ &\ \  0.60 \ \ &\ \  0.60 \ \ &\ \  16.82 \ \ &\ \  0.82 \\
0.68 \ \ &\ \  0.43 \ \ &\ \  0.43 \ \ &\ \  0.82 \ \ &\ \  16.82
\end{array}
\right).
\end{equation}

\subsection{Sr$_2$VO$_4$}

\begin{equation}
U^{\mathrm{scr}}_{t_{2g}} = \left(
\begin{array}{ccccc}
3.46 \ \ &\ \  2.41 \ \ &\ \  2.41 \\
2.41 \ \ &\ \  3.26 \ \ &\ \  2.36 \\
2.41 \ \ &\ \  2.36 \ \ &\ \  3.26
\end{array}
\right),
\end{equation}

\begin{equation}
U^{\mathrm{bare}}_{t_{2g}} = \left(
\begin{array}{ccccc}
15.91 \ \ &\ \  14.28 \ \ &\ \  14.28 \\
14.28 \ \ &\ \  15.18 \ \ &\ \  13.96 \\
14.28 \ \ &\ \  13.96 \ \ &\ \  15.18
\end{array}
\right),
\end{equation}

\begin{equation}
J^{\mathrm{scr}}_{t_{2g}} = \left(
\begin{array}{ccccc}
3.46 \ \ &\ \  0.45 \ \ &\ \  0.45 \\
0.45 \ \ &\ \  3.26 \ \ &\ \  0.43 \\
0.45 \ \ &\ \  0.43 \ \ &\ \  3.26
\end{array}
\right),
\end{equation}

\begin{equation}
J^{\mathrm{bare}}_{t_{2g}} = \left(
\begin{array}{ccccc}
15.91 \ \ &\ \  0.59 \ \ &\ \  0.59 \\
0.59 \ \ &\ \  15.18 \ \ &\ \  0.57 \\
0.59 \ \ &\ \  0.57 \ \ &\ \  15.18
\end{array}
\right).
\end{equation}

For the $d$ model in Sr$_2$VO$_4$, the orbital index runs as ($d_{xy}$, $d_{yz}$, $d_{xz}$, $d_{y^2-z^2}$, $d_{3x^2-r^2}$),
in order to compare its effective interaction parameters with those for Sr$_2$VO$_3$H.

\begin{equation}
U^{\mathrm{scr}}_{d} = \left(
\begin{array}{ccccc}
3.48 \ \ &\ \ 2.42 \ \ &\ \  2.42 \ \ &\ \  2.44 \ \ &\ \  2.71\\
2.42 \ \ &\ \ 3.27 \ \ &\ \  2.36 \ \ &\ \  2.67 \ \ &\ \  2.32\\
2.42 \ \ &\ \ 2.36 \ \ &\ \  3.27 \ \ &\ \  2.38 \ \ &\ \  2.62\\
2.44 \ \ &\ \ 2.67 \ \ &\ \  2.38 \ \ &\ \  3.33 \ \ &\ \  2.36\\
2.71 \ \ &\ \ 2.32 \ \ &\ \  2.62 \ \ &\ \  2.36 \ \ &\ \  3.49
\end{array}
\right),
\end{equation}

\begin{equation}
U^{\mathrm{bare}}_{d} = \left(
\begin{array}{ccccc}
15.95 \ \ &\ \ 14.31 \ \ &\ \  14.31 \ \ &\ \  14.22 \ \ &\ \  15.04\\
14.31 \ \ &\ \ 15.19 \ \ &\ \  13.97 \ \ &\ \  14.47 \ \ &\ \  14.09\\
14.31 \ \ &\ \ 13.97 \ \ &\ \  15.19 \ \ &\ \  13.89 \ \ &\ \  14.66\\
14.22 \ \ &\ \ 14.47 \ \ &\ \  13.89 \ \ &\ \  15.13 \ \ &\ \  14.04\\
15.04 \ \ &\ \ 14.09 \ \ &\ \  14.66 \ \ &\ \  14.04 \ \ &\ \  15.98
\end{array}
\right),
\end{equation}

\begin{equation}
J^{\mathrm{scr}}_{d} = \left(
\begin{array}{ccccc}
3.48 \ \ &\ \ 0.45 \ \ &\ \  0.45 \ \ &\ \ 0.44 \ \ &\ \  0.37\\
0.45 \ \ &\ \ 3.27 \ \ &\ \  0.44 \ \ &\ \  0.30 \ \ &\ \  0.49\\
0.45 \ \ &\ \ 0.44 \ \ &\ \  3.27 \ \ &\ \  0.43 \ \ &\ \  0.36\\
0.44 \ \ &\ \ 0.30 \ \ &\ \  0.43 \ \ &\ \  3.33 \ \ &\ \  0.54\\
0.37 \ \ &\ \ 0.49 \ \ &\ \  0.36 \ \ &\ \  0.54 \ \ &\ \  3.49
\end{array}
\right),
\end{equation}

\begin{equation}
J^{\mathrm{bare}}_{d} = \left(
\begin{array}{ccccc}
15.95 \ \ &\ \ 0.59 \ \ &\ \  0.59 \ \ &\ \  0.57 \ \ &\ \  0.43\\
0.59 \ \ &\ \ 15.19 \ \ &\ \  0.57 \ \ &\ \  0.33 \ \ &\ \  0.67\\
0.59 \ \ &\ \ 0.57 \ \ &\ \  15.19 \ \ &\ \  0.57 \ \ &\ \  0.42\\
0.57 \ \ &\ \ 0.33 \ \ &\ \  0.57 \ \ &\ \  15.13 \ \ &\ \  0.79\\
0.43 \ \ &\ \ 0.67 \ \ &\ \  0.42 \ \ &\ \ 0.79 \ \ &\ \  15.98
\end{array}
\right).
\end{equation}

\subsection{Sr$_2$VO$_3$H}

\begin{equation}
U^{\mathrm{scr}}_{t_{2g}} = \left(
\begin{array}{ccccc}
2.51 \ \ &\ \  1.78 \ \ &\ \  1.63 \\
1.78 \ \ &\ \  2.84 \ \ &\ \  1.78 \\
1.63 \ \ &\ \  1.78 \ \ &\ \  2.46
\end{array}
\right),
\end{equation}

\begin{equation}
U^{\mathrm{bare}}_{t_{2g}} = \left(
\begin{array}{ccccc}
14.93\ \ &\ \  13.91 \ \ &\ \  13.38 \\
13.91\ \ &\ \  15.25 \ \ &\ \  13.59 \\
13.38\ \ &\ \  13.59 \ \ &\ \  14.26
\end{array}
\right),
\end{equation}

\begin{equation}
J^{\mathrm{scr}}_{t_{2g}} = \left(
\begin{array}{ccccc}
2.51 \ \ &\ \  0.43 \ \ &\ \  0.39 \\
0.43 \ \ &\ \  2.84 \ \ &\ \  0.41 \\
0.39 \ \ &\ \  0.41 \ \ &\ \  2.46
\end{array}
\right),
\end{equation}

\begin{equation}
J^{\mathrm{bare}}_{t_{2g}} = \left(
\begin{array}{ccccc}
14.93 \ \ &\ \  0.57 \ \ &\ \  0.53 \\
0.57 \ \ &\ \  15.25 \ \ &\ \  0.55 \\
0.53 \ \ &\ \  0.55 \ \ &\ \  14.26
\end{array}
\right).
\end{equation}

For the $d$ model in Sr$_2$VO$_3$H, the orbital index runs as ($d_{xy}$, $d_{yz}$, $d_{xz}$, $d_{y^2-z^2}$, $d_{3x^2-r^2}$),
where hydrogen atoms are aligned along the $x$ direction.

\begin{equation}
U^{\mathrm{scr}}_{d} = \left(
\begin{array}{ccccc}
3.60 \ \ &\ \ 2.72 \ \ &\ \  2.58 \ \ &\ \ 2.75 \ \ &\ \  2.67\\
2.72 \ \ &\ \ 3.70 \ \ &\ \  2.64 \ \ &\ \ 3.11 \ \ &\ \  2.42\\
2.58 \ \ &\ \ 2.64 \ \ &\ \  3.36 \ \ &\ \ 2.68 \ \ &\ \  2.58\\
2.75 \ \ &\ \ 3.11 \ \ &\ \  2.68 \ \ &\ \  3.86 \ \ &\ \  2.47\\
2.67 \ \ &\ \ 2.42 \ \ &\ \  2.58 \ \ &\ \  2.47 \ \ &\ \  3.14
\end{array}
\right),
\end{equation}

\begin{equation}
U^{\mathrm{bare}}_{d} = \left(
\begin{array}{ccccc}
15.20 \ \ &\ \ 14.14 \ \ &\ \  13.49 \ \ &\ \  14.32 \ \ &\ \  13.48\\
14.14 \ \ &\ \ 15.49 \ \ &\ \  13.68 \ \ &\ \  15.03 \ \ &\ \  13.12\\
13.49 \ \ &\ \ 13.68 \ \ &\ \  14.25 \ \ &\ \  13.88 \ \ &\ \  13.05\\
14.32\ \ &\ \ 15.03\ \ &\ \  13.88 \ \ &\ \ 16.04 \ \ &\ \  13.31\\
13.48 \ \ &\ \ 13.12 \ \ &\ \  13.05 \ \ &\ \  13.31 \ \ &\ \  13.56
\end{array}
\right),
\end{equation}

\begin{equation}
J^{\mathrm{scr}}_{d} = \left(
\begin{array}{ccccc}
3.60 \ \ &\ \ 0.46 \ \ &\ \  0.42 \ \ &\ \  0.43 \ \ &\ \  0.31\\
0.46 \ \ &\ \ 3.70 \ \ &\ \  0.43 \ \ &\ \  0.32 \ \ &\ \  0.45\\
0.42 \ \ &\ \ 0.43 \ \ &\ \  3.36 \ \ &\ \  0.41 \ \ &\ \  0.30\\
0.43 \ \ &\ \ 0.32 \ \ &\ \  0.41 \ \ &\ \  3.86 \ \ &\ \  0.49\\
0.31 \ \ &\ \ 0.45 \ \ &\ \  0.30 \ \ &\ \  0.49 \ \ &\ \  3.14
\end{array}
\right),
\end{equation}

\begin{equation}
J^{\mathrm{bare}}_{d} = \left(
\begin{array}{ccccc}
15.20 \ \ &\ \ 0.58 \ \ &\ \  0.54 \ \ &\ \  0.57 \ \ &\ \  0.37\\
0.58 \ \ &\ \ 15.49 \ \ &\ \  0.56 \ \ &\ \  0.34 \ \ &\ \  0.60\\
0.54 \ \ &\ \ 0.56 \ \ &\ \  14.25 \ \ &\ \  0.55 \ \ &\ \  0.36\\
0.57 \ \ &\ \ 0.34 \ \ &\ \  0.55 \ \ &\ \  16.04 \ \ &\ \  0.68\\
0.37 \ \ &\ \ 0.60 \ \ &\ \  0.36 \ \ &\ \  0.68 \ \ &\ \  13.56
\end{array}
\right).
\end{equation}

\end{document}